\documentclass[useAMS,usenatbib]{mn2e}
\usepackage{graphicx}
\usepackage[usenames]{color}
\usepackage{euscript}

\title[Kinematics and stellar populations of IC~3653]{Kinematics and stellar populations of the dwarf elliptical galaxy IC~3653}

\author[I. V. Chilingarian, P. Prugniel, O. K. Sil'chenko, and V. L. Afanasiev]
{I. V. Chilingarian$^{1,2,3}$\thanks{E-mail: chil@sai.msu.su}, P. Prugniel$^{2,4}$, O. K. Sil'chenko$^{1}$, and V. L. Afanasiev$^{5}$\\
$^{1}$Sternberg Astronomical Institute of the Moscow State University,
Universitetsky pr. 13, Moscow, 119992, Russia\\
$^{2}$Universit\'e de Lyon, Lyon, F-69000, France; Universit\'e Lyon 1, Villeurbanne, F-69622, France; 
Centre de Recherche\\ Astronomique de Lyon, Observatoire de Lyon, 9 Av. Charles Andr\'e, 
Saint-Genis Laval, F-69561, France ; CNRS, UMR 5574 ;\\ Ecole Normale Sup\'erieure de Lyon, Lyon, France\\
$^{3}$LERMA Observatoire de Paris-Meudon, 61 Av. de l'Observatoire, Paris, 75014, France\\
$^{4}$GEPI Observatoire de Paris-Meudon, 5 place Jules Janssen, Meudon, 92195, France\\
$^{5}$Special Astrophysical Observatory of the Russian Academy of Sciences,
Nizhniy Arkhyz, Karachaevo-Cherkesia, 369167, Russia\\}
\begin{document}

\date{Accepted 2007 January 24. Received 2007 January 15; in original form 2006 August 03}

\pagerange{\pageref{firstpage}--\pageref{lastpage}} \pubyear{2007}

\maketitle

\label{firstpage}

\begin{abstract}
We present the first 3D observations of a diffuse elliptical galaxy (dE).
The good quality data (S/N up to 40) reveal the kinematical signature
of an embedded stellar disc, reminiscent of what is commonly observed
in elliptical galaxies, though similarity of their origins is questionable.
Colour map built from HST ACS images confirms the
presence of this disc. Its characteristic scale (about 3~arcsec = 250~pc) is
about a half of galaxy's effective radius, and its metallicity is 0.1--0.2 dex
larger than the underlying population.
Fitting the spectra with synthetic single stellar populations (SSP) we
found an SSP-equivalent age of 5~Gyr and nearly solar metallicity
[Fe/H]=-0.06~dex. We checked that these determinations are consistent 
with those based on Lick indices, but have smaller error bars. 
The kinematical discovery of a stellar disc in
dE gives additional support to an evolutionary link from dwarf irregular
galaxies due to stripping of the gas against the intra-cluster medium.

\end{abstract}


\section{Introduction}
\label{secintro}

Giant elliptical galaxies (E) were believed to be simple objects,
rotationally supported and featureless until the kinematical observations of
NGC~4697 by Bertola \& Capaccioli (1975) made realize that at least some of
them were supported by anisotropic velocity dispersions (Binney 1976,
Illingworth 1977). In parallel, new imaging and processing techniques
revealed the presence of dust lanes (Bertola \& Galletta 1978, Sadler \&
Gerhard 1985) and fine structures (e. g. shells, Malin \& Carter 1983) in
many E's.

Diffuse elliptical galaxies (dE, also called dwarf elliptical or dwarf
spheroidal galaxies) followed the same line more recently when high quality
images revealed 
fine structures, in particular presence
of a spectacular and intriguing stellar spiral in IC~3328 (Jerjen et al.
2000) and IC~783 (Barazza et al. 2002) or broader structures, like embedded
discs or bars in a number of dE's (Barazza et al. 2002). Evidences for
ubiquity of discs in bright dE galaxies in the Virgo cluster, based on
multicolour photometry, have been shown by Lisker et al. (2006)
A number of recent
papers presented also long slit spectroscopic data (Simien \& Prugniel 2002,
Pedraz et al. 2001, Geha et al. 2002, 2003, de Rijcke et al. 2001, van Zee et al.
2004a,b) revealing a diversity of degree of rotational support and even some
complex structures (de Rijcke et al. 2004, Thomas et al. 2006). The stellar
population may also be inhomogeneous (Michielsen et al. 2003) and an ionized ISM has been
detected in some objects (Michielsen et al. 2004). 

The complexity of elliptical galaxies put severe constraints on the
scenario of their formation and evolution. Ellipticals are thought to form
at high redshifts from collapse and hierarchical mergers and accidentally
lately evolve through major mergers. For what concerns dE's, the
main classes of physical mechanisms intervening in the formation and
evolution are (1) the feedback of the star formation on the interstellar
medium and (2) the environment.

The stellar population of most present dE's may not have formed in large
galaxies, because the metallicity, [Fe/H]$\approx$-0.3, (Geha et al., 2003a, Van
Zee et al. 2004b), is typically smaller than that of large bulges or E's.
It is thought to have formed in small galaxies where the supernova driven
winds certainly play an important role in controlling star formation rate
and metal enrichment. In the smallest galaxies the major part of gas and
produced metals will be spread to the intergalactic medium. However duration
of star formation should be long enough, because recent studies (Geha et al.,
2003, Van Zee et al. 2004b) demonstrate that dE galaxies exhibit solar [Mg/Fe],
while in case of short star formation episode they would have been
iron-deficient.

The environment also obviously plays a major role through three phenomena:
\begin{enumerate}
\item the ram pressure stripping against the intra-cluster medium (Marcolini et
al. 2003),
\item the tidal harassment due to distant and repeated encounters with
other cluster members (Moore et al. 1998) and
\item the collisions involving large gas rich objects that may disrupt gas
clouds from which a dwarf galaxy may born (tidal dwarfs) and evolve
into a dE after fading of the young stellar population (Duc \& Mirabel 1999).
\end{enumerate}

The latter possibility may probably not be the main scenario because dE's
are one order of magnitude more numerous than large galaxies which in
average have suffered one major collision in their lifetime, while the merger
remnants in the local Universe are only accompanied by a few tidal dwarfs
candidates which have generally much smaller mass than bright ellipticals
(Weilbacher et al. 2000).

The two first mechanisms are undoubtedly the major ingredients. Ram pressure
stripping is an accepted explanation for the HI depletion of spiral
galaxies in clusters and any type of gas rich galaxies will experience it.
It may well explain the relation between the morphology of dwarfs and the
density of the environment (Binggeli et al. 1990, van den Bergh 1994):
low-density environment is mostly populated by dIrr's, while dE's reside
mostly in clusters where ram pressure stripping is expected to be the most
efficient. Recently, Conselice et al. (2003) discovered HI in two dE's in the
Virgo cluster, which they interpret as transition objects on their way in
the morphological transformation from dIrr to dE. 
Conselice et al. (2001) argue that dE's are recently accreted galaxies
which morphologically evolve from late type galaxies when they cross the
center of the cluster.

There is yet no decisive test to disentangle the roles of each mechanism.
In this paper we present 3D spectroscopic observations in order to
bring further observational constraints. Velocity fields and spatial
distribution of the stellar population are most needed to check if
counterparts of the observed kinematical sub-structures can be detected.

We are presenting here the first 3D observations of a dE. IC~3653
is a bright dE galaxy belonging to the Virgo cluster (Binggeli et al.
1985). In Table~\ref{tabic3653params} we summarize its main characteristics.
IC~3653 was chosen because it
is amongst the most luminous dE in Virgo and has a
relatively high surface brightness. It is located 2.7~deg from the center
of the cluster, i. e. 0.8~Mpc in projected distance. Its radial velocity 588$\pm4$~km~s$^{-1}$
(this work) confirms its membership to the Virgo cluster, the velocity difference from the
mean velocity of Virgo (1054~km~s$^{-1}$, HyperLeda, Paturel et al. 2003 
\footnote{http://leda.univ-lyon1.fr/}) is nearly -470~km~s$^{-1}$. 
IC~3653 is located some 100~kpc in the projected distance from NGC~4621, a
giant elliptical galaxy having a similar radial velocity value
(410~km~s$^{-1}$, HyperLeda) With other low luminosity Virgo cluster
members, in particular IC~809, IC~3652 for which the radial velocities have been
measured, they may belong to a physical substructure of Virgo, crossing the
cluster at 500 km~s$^{-1}$.

Velocity and velocity dispersion
profiles from by Simien \& Prugniel (2002) show some rotation. ACS/HST
archival images from the Virgo cluster ACS survey (C\^ot\'e et al.2004) 
are also available and will be discussed here.

In Sect.~2 we describe the observations and the data reduction. In Sect.~3 we
derive the age and the metallicity using spectrophotometric indices and in
Sect.~4 we introduce and use a new method for measuring the internal kinematics
and the population parameters by fitting directly the spectra against high
spectral resolution population models. The analysis of the ACS images is
presented in Sect.~5. Sect.~6 is the discussion.

\begin{table}
  \begin{tabular}{l l}
\hline
Name & IC3653, VCC1871 \\
Position & J124115.74+112314.0 \\
B & 14.55\\
Distance modulus & 31.15\\
A(B) & 0.13\\
M(B)$_{corr}$ & -16.78\\
Spatial scale & 82 pc~arcsec$^{-1}$ \\
Effective radius, $R_e$ & 6.7 arcsec $\equiv$ 550 pc \\
$\mu_B$, mag~arcsec$^{-2}$ & 20.77 \\
Ellipticity, $\epsilon$ & 0.12\\
S\'ersic exponent, $n$ & 1.2\\
Heliocentric cz, km~s$^{-1}$ & 588 $\pm$ 4 \\
$\sigma_{cent}$, km~s$^{-1}$ & 80 $\pm$ 3 \\
$V_{max}$, km~s$^{-1}$  & 18 $\pm$ 2\\
$V_{max}/\sigma$ & 0.27$\pm$0.08\\
$t$, Gyr (lum. weighted) & 5.2$\pm$0.2\\
$[Z/\mbox{H}]$, dex (lum. weighted) & -0.06$\pm$0.02\\
\hline
  \end{tabular}
  \caption{General characteristics of IC~3653. S\'ersic exponent, kinematical
  and stellar population parameters are obtained in this paper, other properties
  are taken from HyperLEDA and Goldmine (Gavazzi et al. 2003, http://goldmine.mib.infn.it/) 
  databases, and from Ferrarese et al. 2006.
  Uncertainties given for age and metallicity correspond to the measurements
  on co-added spectra.
  \label{tabic3653params}
}
\end{table}

\section[]{Spectroscopic observations and data reduction}
\label{secobs}
The spectral data we analyse were obtained with the MPFS integral-field
spectrograph. 

The Multi-Pupil Fiber Spectrograph (MPFS), operated on the 6-m telescope
Bolshoi Teleskop Al'tazimutal'nij (BTA) of the Special Astrophysical Observatory
of the Russian Academy of Sciences, is a fibre-lens spectrograph with a
microlens raster containing $16 \times 16$ square spatial elements together
with 17 additional fibres transmitting the sky background light, taken
four arcminutes away from the object. The size of each element is 1''$
\times$1''. We used the grating 1200 gr mm$^{-1}$ providing the reciprocal
dispersion of 0.75~\AA~pixel$^{-1}$ with a EEV CCD42-40 detector in the 
spectral range 4100\AA -- 5650\AA.

Observations of IC~3653 were made on 2004 May 24 under good
atmosphere conditions (seeing FWHM=1.4''). The total integration time was 2
hours. The
spectral resolution $R=\lambda / \Delta \lambda$, where $\Delta \lambda$ is the
FWHM resolution (width of the line-spread function), as determined by analysing
twilight spectra, varied from $R=1300$ to $R=2200$ (between 2.5\AA\ and 3.3\AA) over
the field of view and wavelength. $R$ is lower in the centre of the field and
increases toward top and bottom and also in the red end of the wavelength range
(Moiseev, 2001).

The following calibration frames were taken during the observations
of IC~3653 with MPFS:
\begin{enumerate}
\item BIAS, DARK.
\item "Etalon": 17 night-sky fibres illuminated by the incandescent bulb.
This frames are used to determine positions of spectra on the frame.
\item "Neon"\ (arc lines): by exposing the spectral lamp filled with
Ar-Ne-He to perform a wavelength calibration.
\item The internal flat field lamp.
\item A spectrophotometric standard ($Feige~56$ for our observations), used
to turn the spectra into absolute flux units.
\item A standard for Lick indices and radial velocity ($HD~137522$ and 
$HD~175743$), used also to measure instrumental response: asymmetry and width of
the line-spread function.
\item "SunSky": twilight sky spectra for additional corrections of the 
systematic errors of the dispersion relation and transparency
differences over the fibres.
\end{enumerate}

\subsection{Data reduction}
\label{subsecdatared}
We use the original IDL software package created
and maintained by one of us (VA), that we modified for this work:
error frames are created using photon statistics and then processed through
all the stages to have realistic error estimates for the fluxes in the
resulting spectrum.

The primary reduction process (up-to obtaining flux-calibrated data cube)
consists of:
\begin{enumerate}
\item Bias subtraction, cosmic ray cleaning.
   Cosmic ray cleaning implies the presence of several frames. Then they are
   normalized and combined into the cube
   (x,y,Num). The cube is then analysed in each pixel through "Num"\ frames.
   All counts exceeding some level (5-$\sigma$) are replaced with the robust
   mean through the column. Then the individual cleaned frames are summed. This
   procedure assumes that all the frames have the same atmospheric conditions
   and cleans only the brighter spikes; though, it was found sufficient for our
   purpose.
\item Creation of the traces of spectra in the "etalon"\ image.
   Accuracy of the traces is usually about 0.02 or 0.03 pixels.
\item Flat field reduction and diffuse light subtraction.
   Flat field is applied to the CCD frames before extracting
   the spectra. The scattered light model is also constructed and
   subtracted from the frames during this step. It is made using parts
   of the frames not covered by spectra and then interpolated with
   low-order polynomials.
\item Creation of the traces for every fibre.
   On this step the traces are determined for each fibre in the
   microlens block (presently 256 fibres) using the night sky fibre 
   traces created on the 2-nd step and interpolation between them
   using the tabulated fibre positions.
\item Spectra extraction.
   Using the fibre traces determined in the previous steps, spectra
   are extracted from science and calibration frames using fixed-width
   Gaussian (usually with FWHM=5~px for the present configuration of the
   spectrograph). The night sky spectra are also extracted from the science
   frames.
\item Creation of dispersion relations.
   Spectral lines in the arc lines frame are identified and dispersion relations
   are computed independently for every fibre.
\item Wavelength rebinning.
   All the spectra of night sky, object and standard stars are rebinned
   independently into logarithm of wavelength (as required for the kinematical 
   analysis). The sampling on the CCD varies between 0.65 and 0.85~\AA\ (FWHM
   resolution 3 to 5 pixels) and we rebinned to a step of 40~km s$^{-1}$, i.
   e. 0.55 to 0.75~\AA, corresponding to the mean oversampling factor 1.2 (we
   checked that this oversampling was high enough to have no measurable effect
   on the result of the analysis).
\item Sky subtraction. The sky spectrum, computed as the median of the 17 night
   sky fibres, is rescaled using flat field to account for the twice larger
   aperture of the night fibres compared to object fibres. Then it is subtracted
   from each fibre.
\item Determination of the spectral sensitivity and absolute
   flux calibration. Using the spectrophotometric standard
   star, the ratio between counts and absolute flux is calculated
   and then approximated with a high-order polynomial function 
   over the whole wavelength range. Finally, the values in
   the data cube are converted into $F_{\lambda}
   [erg \cdot cm^{-2} \cdot s^{-1} \cdot $\AA$^{-1}]$.
\end{enumerate}

\subsection{Spatial adaptive binning}
In our data, the surface brightness, $\mu_{B}$, changes from
18.5~mag~arcsec$^{-2}$ in the centre down to 21.2~mag~arcsec$^{-2}$ in the outer
parts of the field of view. Though the signal-to-noise ratio at the central part
is high enough ($\sim 30$) for detailed
analysis, the accuracy of kinematics and Lick indices measurements in individual
decreases dramatically at the largest radii. To increase the signal-to-noise
ratio in the outer parts, without degrading the resolution in
the centre, we applied the Voronoi adaptive binning procedure proposed by
Cappellari \& Copin (2003) for this exact purpose. It uses variable bin size to
achieve equal signal-to-noise ratio over the field of view. The Voronoi 
2D-binning produces a set of 1D spectra that are then analysed independently.
For the kinematical analysis we will use a target signal-to-noise ratio
of 15, and for stellar population analysis we will use 30.

Besides we will also use a tessellation of the dataset containing only three
bins (3-points binning hereafter): central condensation (3 by 3 arcsec region
around the centre of the galaxy), elongated disky substructure (14 by 7 arcsec)
oriented according to kinematics (see section 4, Fig.~\ref{spec4}, illustrating
locations of bins and demonstrating spectra integrated in them) with the central
region excluded, and the rest of the galaxy. The location of these three bins is
shown in Fig.~\ref{spec4}, and their characteristics in Table~\ref{tabpar3bin}.
Such a physically-stipulated tessellation allows to gain high signal-to-noise
ratios in the bins in order to have high quality estimations of the stellar
population parameters in the regions where they are expected to be homogeneous.

\begin{table}
\begin{tabular}{l c c c c}
Bin & N$_{spax}$ & m(AB) & $\mu$(AB) & S/N \\
\hline
P1 &    9 &  16.3& 18.7& 69\\
P2 &   77 &  15.2& 19.9& 49\\
P3 &  122 &  15.7& 20.9& 21\\
\hline
\end{tabular}
\caption{Parameters of the "3-points"\ binning: number of spatial
elements, mean AB magnitude, mean AB surface 
brightness (mag~arcsec$^{-2}$), and mean signal-to-noise ratio at 5000\AA.
\label{tabpar3bin}
}
\end{table}

\section{SSP age and metallicity derived from Lick indices}
\label{seclick}
A classical and effective method of studying stellar population properties 
exploits diagrams for different pairs of Lick indices (Worthey et al., 1994). 
A grid of values, corresponding
to different ages and metallicities of single stellar population models 
(instantaneous burst, SSP), is plotted
together with the values computed from the observations. 
A proper choice of the pairs of indices, sensitive to mostly age or metallicity
like  H$\beta$ and Mg$b$, allows to determine SSP-equivalent age and 
metallicity.

\begin{figure*}
\hfil
\begin{tabular}{c c}
 (a) & (b) \\
 \includegraphics[width=8cm,height=8cm]{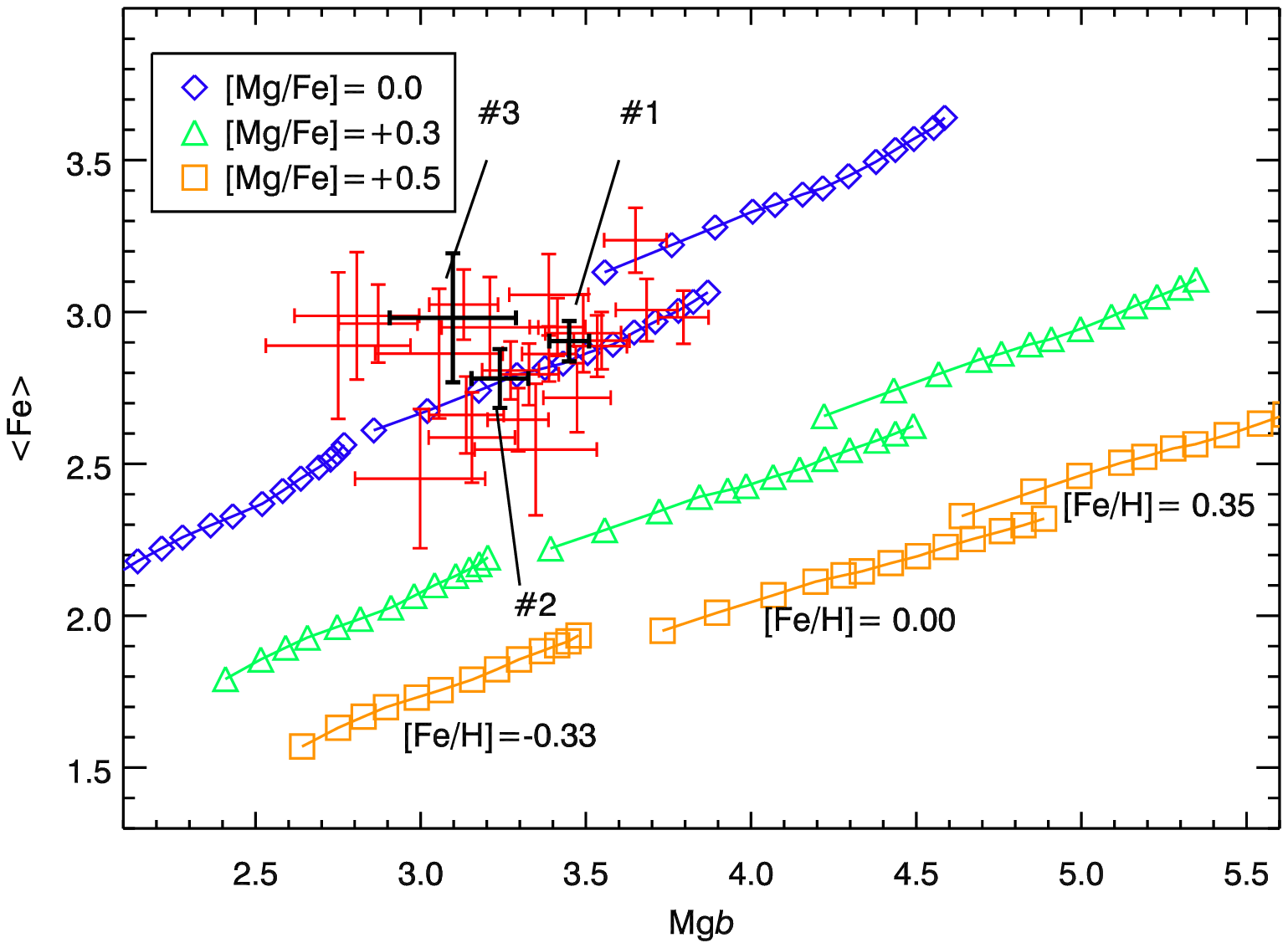} &
 \includegraphics[width=8cm,height=8cm]{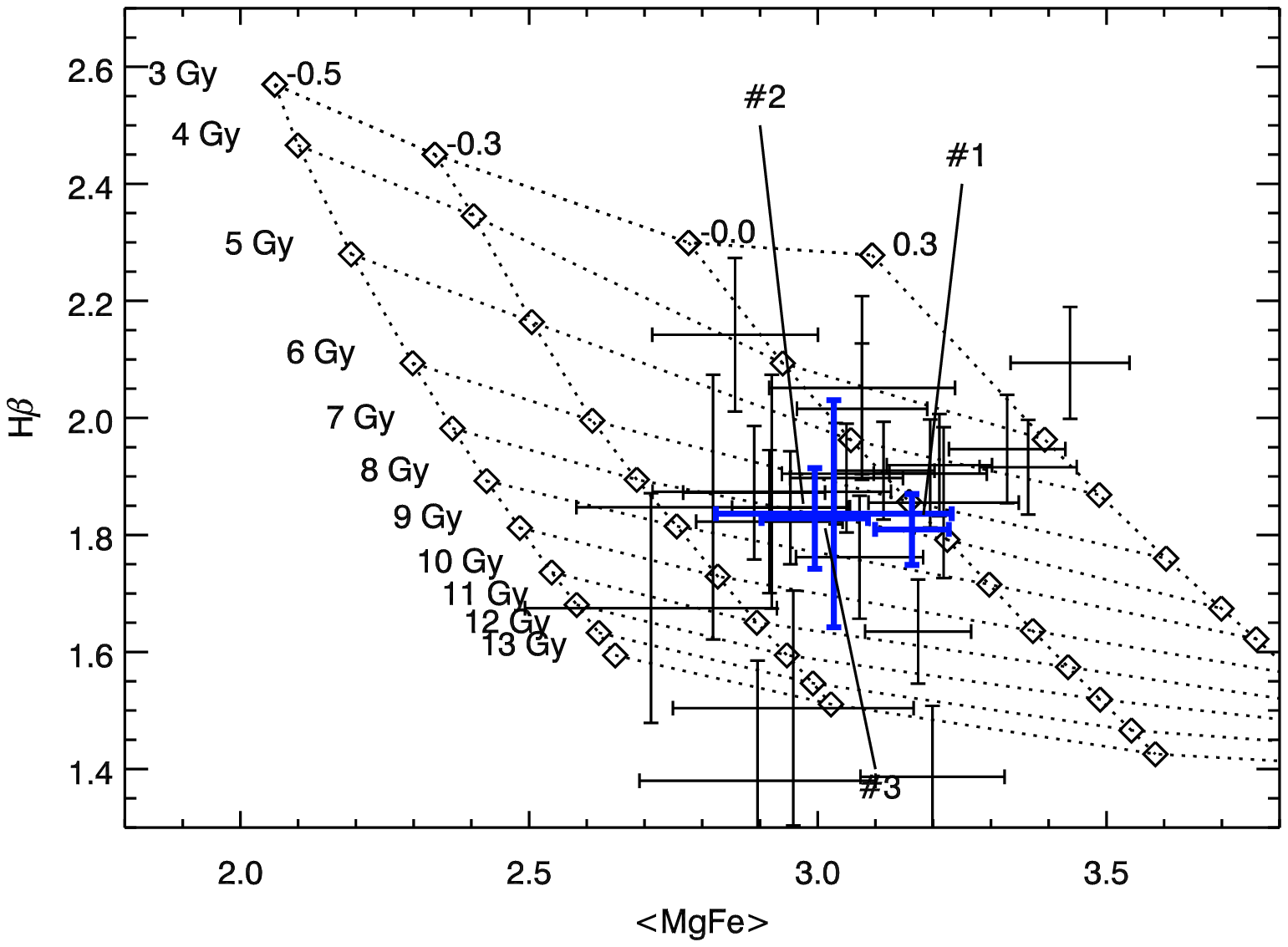} \\
 (c) & (d) \\
 \includegraphics[width=8cm,height=8cm]{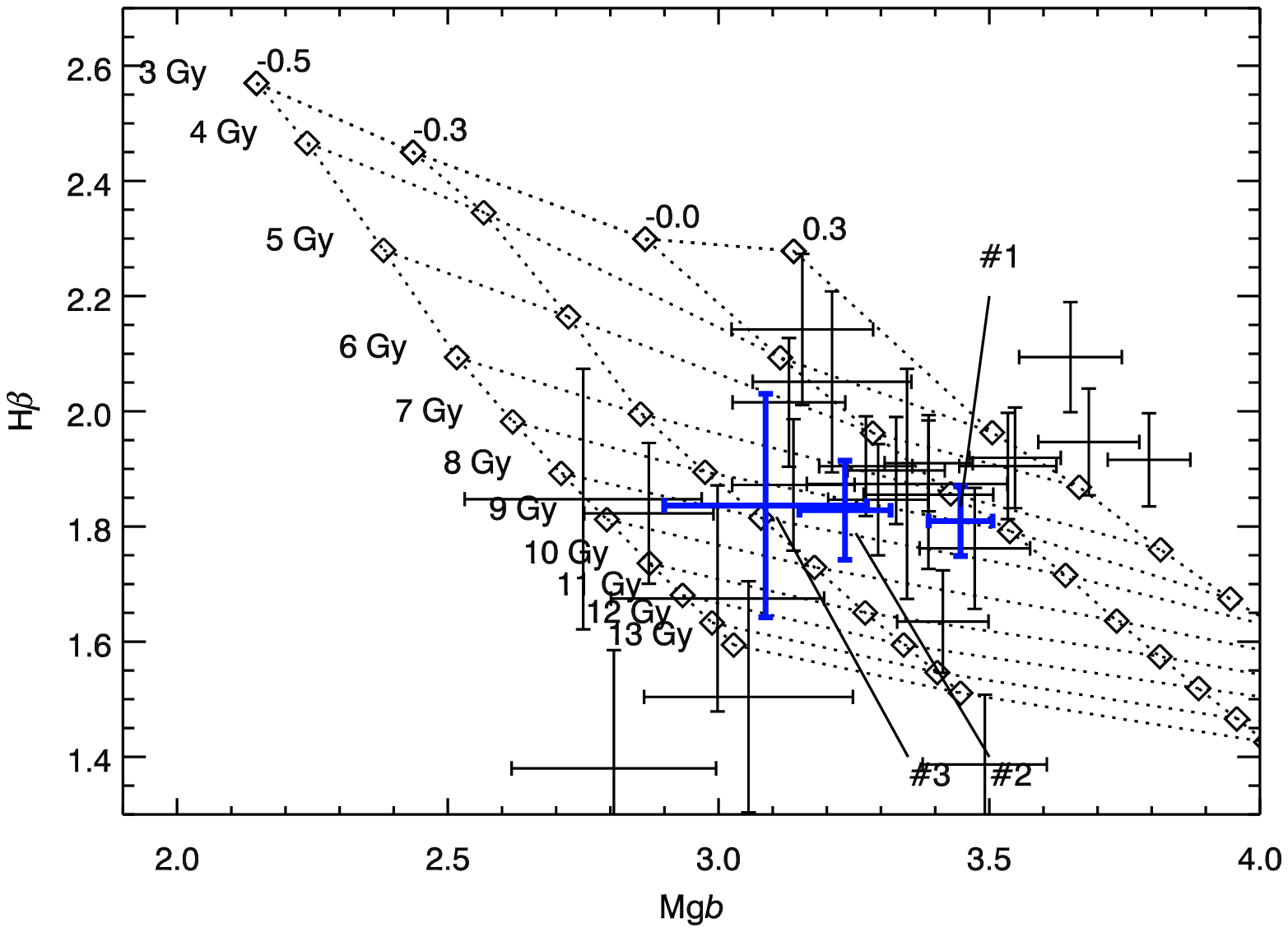} &
 \includegraphics[width=8cm,height=8cm]{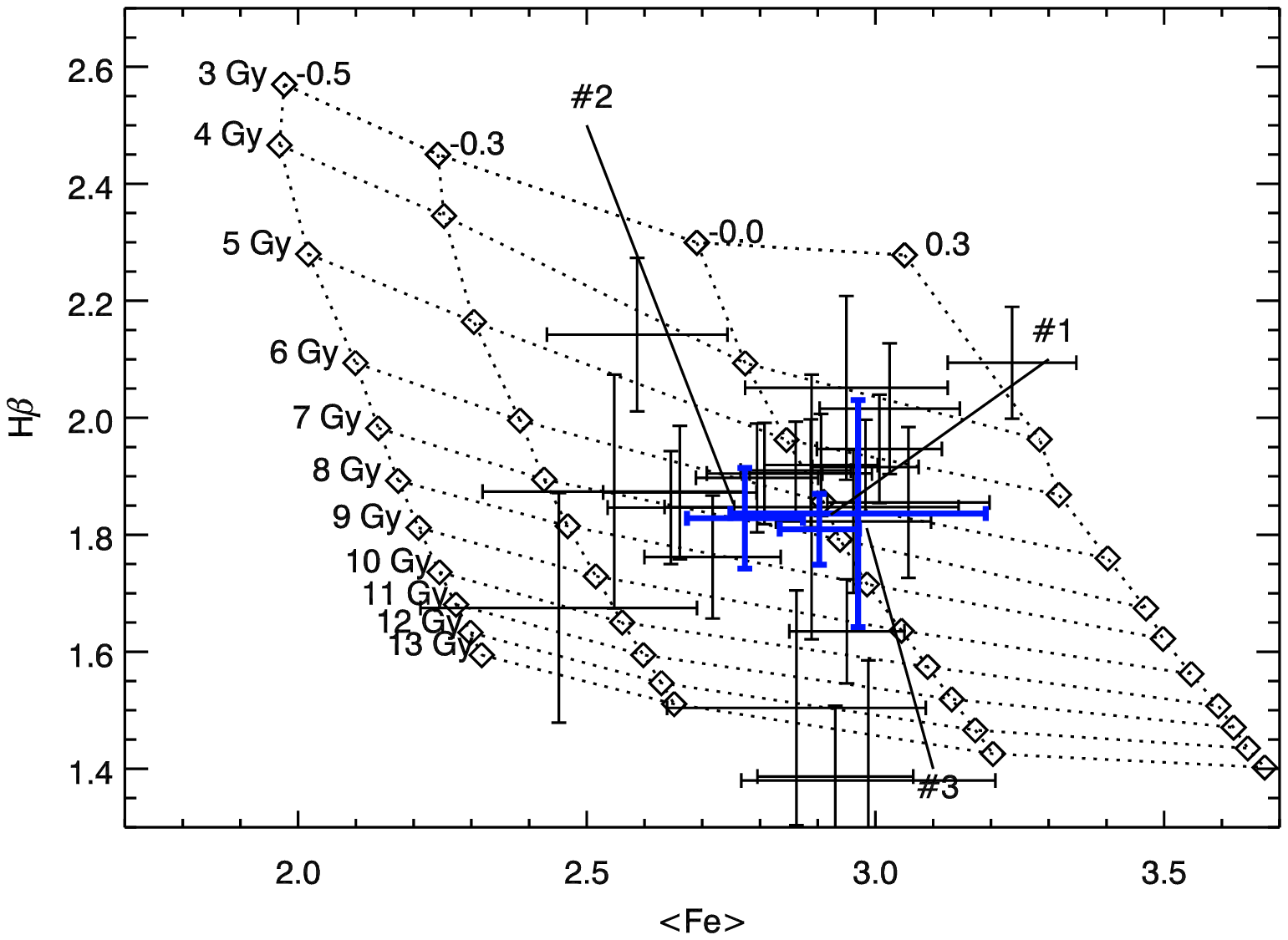} \\
\end{tabular}

\caption{The Mg$b$ - $<$Fe$>'$ (a), H${\beta}$ - [MgFe] (b), H${\beta}$ -
Mg$b$ (c), and H${\beta}$ - $<$Fe$>'$ (d) diagrams. On (a) the models from Thomas
et al. (2003) are plotted. On (b), (c), and (d) the displayed grid is
constructed of the values of the Lick indices for PEGASE.HR synthetic spectra
(SSP) for different ages and metallicities. Bold crosses with pointers represent
measurements for 3 regions of the galaxy (see text), thin crosses are for
individual bins for Voronoi tessellation with target S/N=30.
\label{lickdiag}
} \hfil
\end{figure*}

We use a grid of models computed with the evolutionary synthesis code:
PEGASE.HR (Le Borgne et al., 2004). These models are based on the 
empirical stellar library ELODIE.3 (Prugniel \& Soubiran 2001, 2004)
and are therefore bound to the [Mg/Fe] abundance pattern of the solar 
neighborhood (see Wheeler et al., 1989).
To show that this limitation is not critical for our (low-mass) galaxy,
Fig.~\ref{lickdiag}a presents the Mg$b$ versus $<$Fe$>$ diagram with the models
by Thomas et al. (2003) for different [Mg/Fe] ratios overplotted. These data
allow to conclude that IC~3653 has solar [Mg/Fe] abundance ratio with a
precision of about 0.05~dex.

A critical limitation of Lick indices is their sensitivity
to missed/wrong values in the data, for example due to imperfections of the
detector, or uncleared cosmic ray hits. A simple interpolation of the missed
values is a poor solution, because if some important detail in the
spectrum, e.g. absorption line, is affected, the final measurement of the index
will be biased. In addition, the definition of the indices (see equations 1,
2, and 3 in Worthey et al. 1994) does not allow to flag or to decrease the
weight of low quality values.

Due to a defect of the detector, our data  have
a 3pixel wide bad region (hot pixels) in the middle of the blue continuum
of Mg$b$. So, strictly speaking, we could not measure Mg$b$ at all,
neither H${\beta}$ on a significant part of the field of view.

As a workaround, we replaced all the missing or flagged values in the
data cube by the corresponding values of the best-fitting model determined
as explained in the next section. Therefore, instead of of a mathematical
interpolation of missing values, we are using model predictions.

\begin{table}
\begin{tabular}{l l l l}
Name & bin 1 & bin 2 & bin 3 \\
\hline
Ca4227& 1.062 $\pm$  0.080 & 0.874 $\pm$  0.190 & 0.689 $\pm$  0.700 \\
 & 1.096              & 1.020              & 1.092              \\
G4300& 5.106 $\pm$  0.135 & 5.042 $\pm$  0.313 & 6.923 $\pm$  1.040 \\
 & 4.995              & 4.820              & 4.979              \\
Fe4383& 5.794 $\pm$  0.176 & 4.940 $\pm$  0.389 & 6.251 $\pm$  1.243 \\
 & 4.861              & 4.458              & 4.696              \\
Ca4455& 1.187 $\pm$  0.089 & 1.096 $\pm$  0.186 & 0.750 $\pm$  0.561 \\
 & 1.336              & 1.235              & 1.313              \\
Fe4531& 2.858 $\pm$  0.125 & 2.326 $\pm$  0.260 & 1.595 $\pm$  0.777 \\
 & 3.499              & 3.367              & 3.457              \\
Fe4668& 6.070 $\pm$  0.181 & 5.776 $\pm$  0.361 & 5.685 $\pm$  1.025 \\
 & 5.114              & 4.602              & 4.808              \\
H$\beta$& 1.841 $\pm$  0.067 & 1.823 $\pm$  0.122 & 1.800 $\pm$  0.304 \\
 & 1.908              & 1.966              & 1.878              \\
Fe5015& 5.121 $\pm$  0.138 & 4.767 $\pm$  0.246 & 4.912 $\pm$  0.584 \\
 & 5.491              & 5.211              & 5.292              \\
Mg$b$& 3.575 $\pm$  0.065 & 3.550 $\pm$  0.116 & 3.771 $\pm$  0.278 \\
 & 3.361              & 3.203              & 3.334              \\
Fe5270& 3.016 $\pm$  0.072 & 2.969 $\pm$  0.132 & 3.362 $\pm$  0.309 \\
 & 3.059              & 2.886              & 2.963              \\
Fe5335& 2.708 $\pm$  0.084 & 2.551 $\pm$  0.155 & 2.526 $\pm$  0.367 \\
 & 2.675              & 2.524              & 2.598              \\
Fe5406& 1.687 $\pm$  0.065 & 1.667 $\pm$  0.121 & 1.542 $\pm$  0.286 \\
 & 1.840              & 1.720              & 1.779              \\
$<$Fe$>'$& 2.930 $\pm$  0.075 & 2.852 $\pm$  0.138 & 3.128 $\pm$  0.325 \\
 & 2.952              & 2.785              & 2.861              \\
$[$MgFe$]$ & 3.236 $\pm$  0.070 & 3.182 $\pm$  0.127 & 3.435 $\pm$  0.301 \\
 & 3.150              & 2.987              & 3.089              \\
\hline
\end{tabular}
\caption{Measurements of selected Lick indices for the 3-points binning. All
values are in \AA. The first line for each index corresponds to the measurements
made on the real spectra and the second on the best-fitting optimal templates
(see text). (Note that the Fe indices Fe4531 and Fe5015 are essentially
sensitive to titanium, see Sil'chenko \& Shapovalova, 1989.)
\label{tablick3b}
}
\end{table}

\begin{figure*}
\hfil
\begin{tabular}{c c c c}
 (a) & (b) & (c) & (d) \\
 \includegraphics[width=4cm]{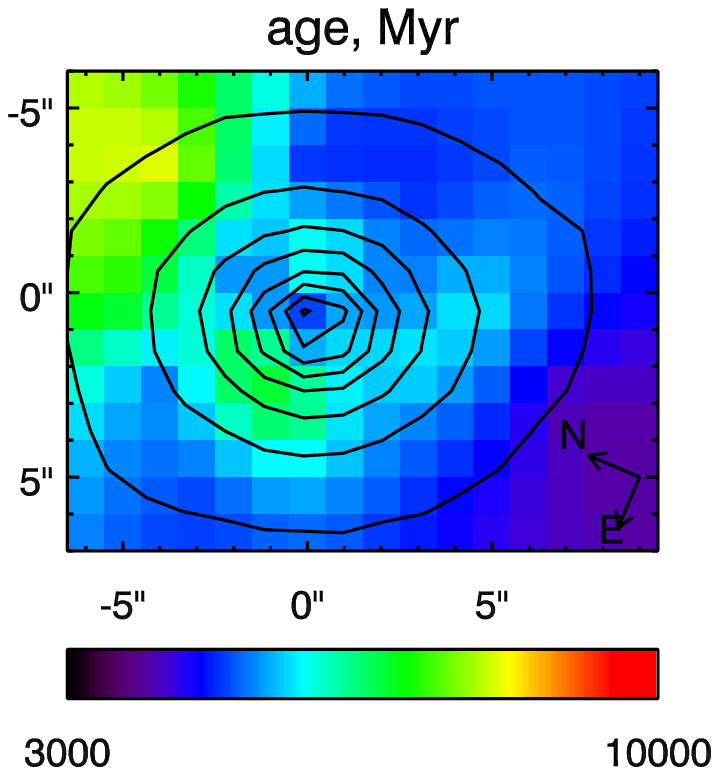} &
 \includegraphics[width=4cm]{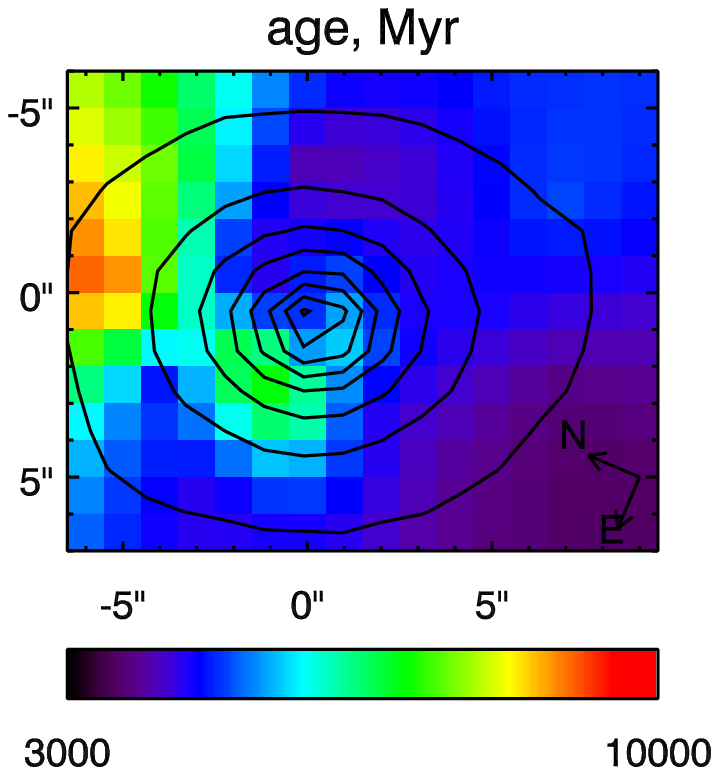} &
 \includegraphics[width=4cm]{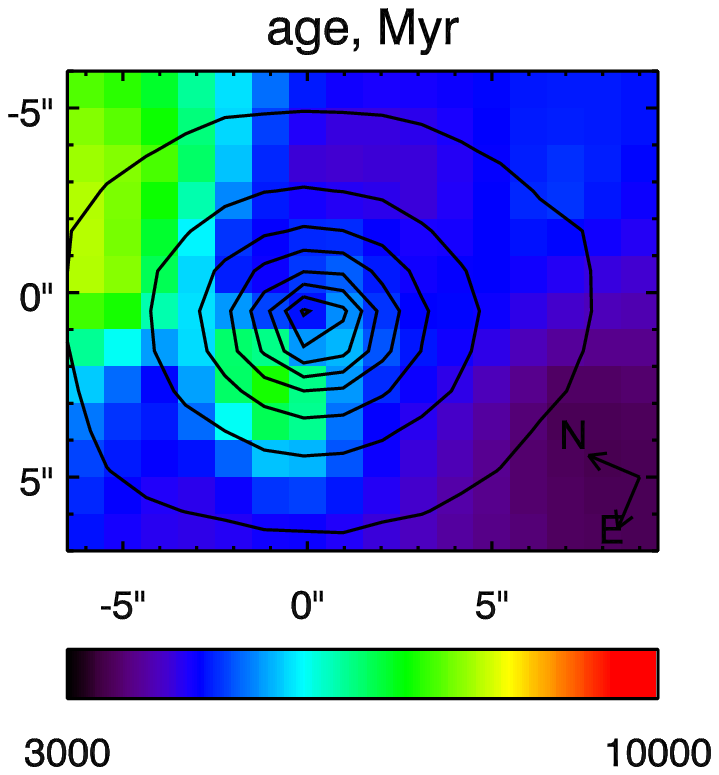} &
 \includegraphics[width=4cm]{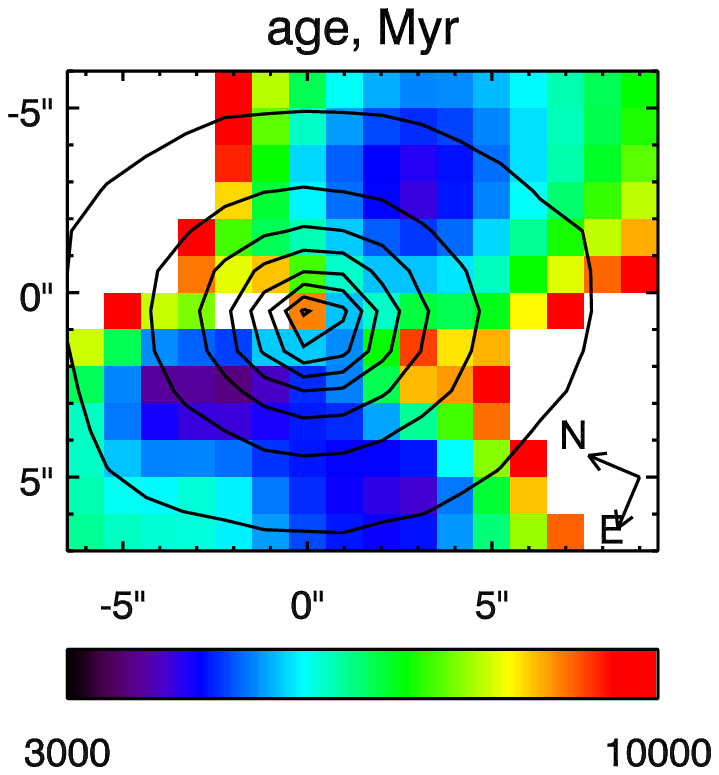} \\
 \includegraphics[width=4cm]{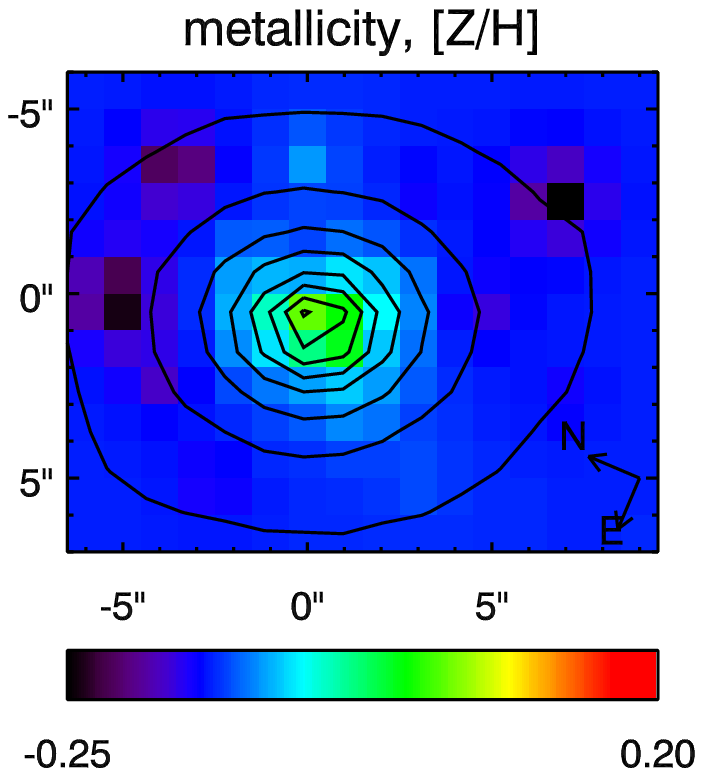} &
 \includegraphics[width=4cm]{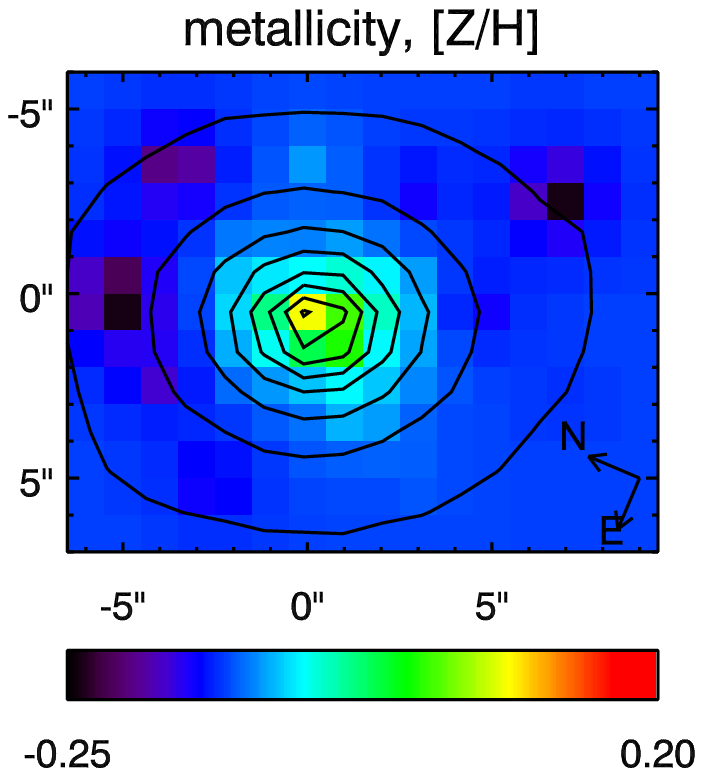} &
 \includegraphics[width=4cm]{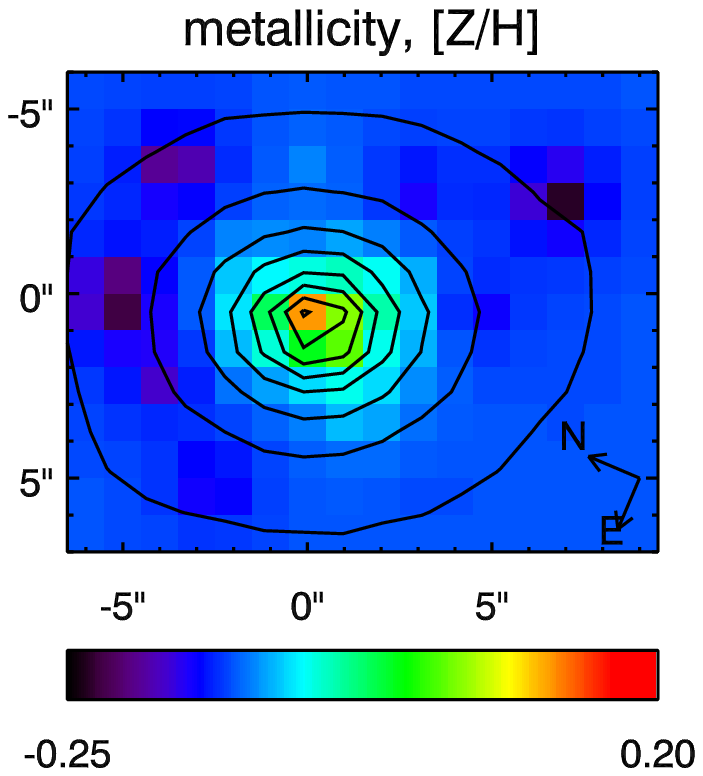} &
 \includegraphics[width=4cm]{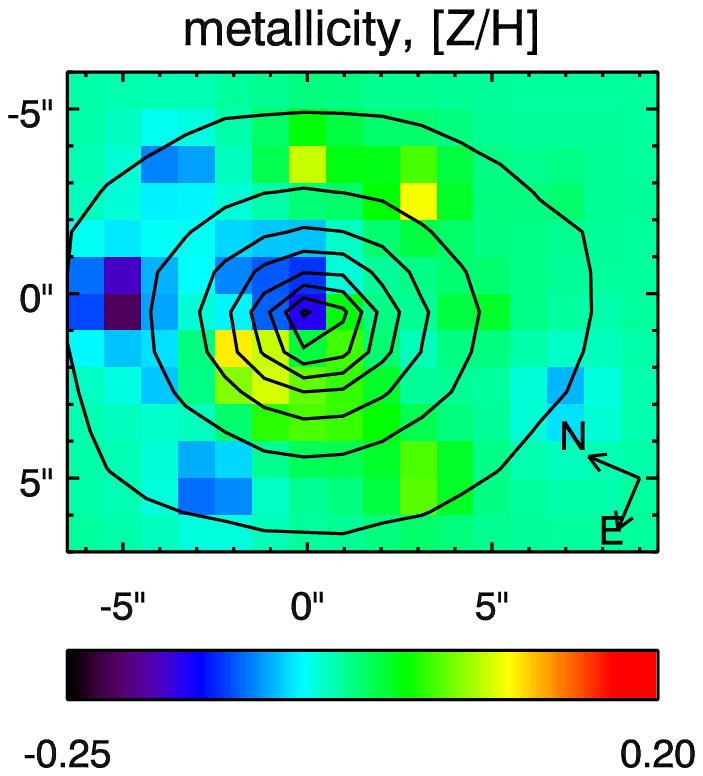} \\
\end{tabular}

\caption{Maps for age and metallicity obtained by inverting bi-index grids for
indices measured on optimal template spectra: (a) Mg$b$ - H${\beta}$, (b) 
[MgFe] - H${\beta}$, (c) $<$Fe$>$ - H${\beta}$; and on real data: 
(d) $<$Fe$>$ - H${\beta}$.
\label{lickfld}
}
\hfil
\end{figure*}

We measured Lick indices as defined in Trager et al. (1998), and we also derived
the combined iron index $<$Fe$>'=0.72$Fe$_{5270} + 0.28$Fe$_{5335}$, and 
"abundance-insensitive"\ [MgFe]$ = \sqrt {\mbox{Mg} <\mbox{Fe}>'}$ (Thomas et al. 2003).
A good intermediate resolution age tracer, H${\gamma}$+Mg+Fe$_{125}$ (Vazdekis
\& Arimoto 1999) cannot be used, because the required signal-to-noise ratio of
about 100 at around $\lambda = 4340$\AA\ cannot be achieved even after
co-adding all the spectra in the data cube due to lower S/N in the blue end
of the spectral range. The statistical errors on the measurements of Lick
indices were computed according to Cardiel et al. (1998). Table~\ref{tablick3b}
presents the measurements of selected Lick indices for the 3-points binning and
Fig.~\ref{lickdiag}c presents the main index-index pairs that are used to
determine age and metallicity.

We see almost no population difference among three bins within the precision we
reach. The age (see Table~\ref{tabcomptz3b}) is around 6~Gyr, metallicity is
about solar for "P1", and slightly subsolar for "P2"\ and "P3".

The large scatter of the measurements, seen on Fig.~\ref{lickdiag},                          
results in a spread of age estimations from 4 to 13~Gyr.                    
It is mainly caused by an imperfect cleaning of the spikes                      
and dark pixels but also weak nebular emission may alter                        
these indices: H$\beta$ index might be affected by emission in H$\beta$,
Mg$b$ -- by [NI] ($\lambda = 5199$\AA) laying in the red continuum
region. Though we do not see any significant emission line                      
residuals when we subtract the best fitting model, we can                       
not exclude completely this effect.

To reduce the scatter, we measured Lick indices on the optimal templates fitted
to the data. This approach may produce biased results in case of model mismatch,
due for example to inconsistent abundance ratios between the models and the real
stellar population. However, we believe it is not the case, since IC~3653
exhibits solar Mg/Fe abundance ratio (see Fig.~\ref{lickdiag}a).

We made inversions of the bi-index grids for three combinations of indices:
Mg$b$ - H${\beta}$, [MgFe] - H${\beta}$, $<$Fe$>'$ - H${\beta}$. The maps shown
in Fig.~\ref{lickfld} represent interpolated values of the parameters between
intensity-weighted centres of the bins.

The metallicity distribution shows slight gradient from -0.15~dex at the periphery
to +0.10 in the very centre (the average error-bar on the metallicity
measurements using [MgFe] - H${\beta}$ is 0.15~dex).

The median value for age using [MgFe] - H${\beta}$ pair is $6 \pm 2.5$~Gyr,
without significant spatial variation. [MgFe] and $<$Fe$>'$ indices are not very
age sensitive, thus the age estimations depend mostly on values of H${\beta}$,
and are almost equal for all three pairs of indices.

\section{Stellar populations and internal kinematics using pixel fitting}
\label{secssp}

Various methods have been developed to determine the star formation and
metal enrichment history (SFH) directly from 
observed spectra (Ocvirk et al. 2003, Moultaka et al. 2004, de Bruyne et al.
2004). The procedure that we are proposing here, population pixel fitting, is
derived from the penalized pixel fitting method developed by Cappellari \&
Emsellem (2004) to determine the line-of-sight velocity distribution (LOSVD).

The observed spectrum is fitted in pixel space against the population model
convolved with a parametric LOSVD. The population model consists of one or
several star bursts, each of them parametrized by some of their characteristics,
typically age and metallicity for a single burst while the other
characteristics, like IMF, remain fixed. A single minimization returns the
parameters of LOSVD and those of the stellar population.

Ideally, we would like to reconstruct SFH, over all the life of the galaxy.
This means, disentangle internal kinematics and distribution in the HR
diagram from the integrated-light spectrum. This problem has been
discussed in several places (e. g. Prugniel et al. 2002, de Bruyne et al.
2004, Ocvirk et al. 2006), it is clearly extremely degenerated and solutions
can be found only if a simplified model is fitted. 

In this paper we discuss only the simplest case of SSP 
characterised by two parameters: age and 
metallicity. We do not discuss complex SFH, because signal-to-noise ratio
of our data is not sufficient.

The $\chi^2$ value (without penalization) is computed
as follows:
\begin{equation}
	\chi^2 = \sum_{N_{\lambda}}\frac{(F_{i}-P_{1p}(T_{i}(t,Z) \otimes
	\mathcal{L}(v,\sigma,h_3,h_4) + P_{2q}) )^2}{\Delta F_{i}^2},
\label{chi2eq}
\end{equation}
where $\mathcal{L}$ is LOSVD; $F_{i}$ and $\Delta F_{i}$ are observed flux and
its uncertainty; $T_{i}$ is the flux from a SSP spectrum, convolved
by the line-spread function of the spectrograph (LSF, 
see next subsection); $P_{1p}$ and $P_{2q}$ are
multiplicative and additive Legendre polynomial of orders $p$ and $q$ for
correcting a continuum; $t$ is age, $Z$ is metallicity, $v$, $\sigma$,
$h_3$. and $h_4$ are radial velocity, velocity dispersion and
Gauss-Hermite coefficients respectively (Van der Marel \& Franx, 1993). Normally we used no additive polynomial
continuum, and 5-th order multiplicative one, and for IC~3653, which has a low
velocity dispersion resulting in insufficient sampling of the LOSVD, we did not
fit $h_3$ and $h_4$. The problem can be partially linearized: in particular, fitting
of additive polynomial continuum, and relative contributions of sub-populations
$T_{i}$ is done linearly on each evaluation of the non-linear functional. Thus
we end up with 10 free parameters:
$t$, $Z$, 6 coefficients for $P_{mult 5}$, $v$, and $\sigma$.

The main technical part of our method is a non-linear minimization procedure
for $\chi^2$ difference between the observed and modeled spectra. The latter
is made by interpolating a grid of SSP spectra computed with PEGASE.HR and
degraded to match the LSF of the observed spectrum. This grid has 25 steps
in age (10~Myr to 20~Gyr) and 10 steps in metallicity ([Fe/H] from -2.5 to
1.0). Because the minimization procedure requires that the derivatives of
the functions are continuous, we used a two-dimensional spline
interpolation. The non-linear minimization is made with the MPFIT package
(by Craig B. Markwardt, NASA
\footnote{http://cow.physics.wisc.edu/\~ craigm/idl/fitting.html}) implementing
constrained variant of Levenberg-Marquardt minimization.

\subsection{Line spread function of the spectrograph}
Before comparing a synthetic spectrum to an observation, it is required
to transform it as if it was observed with the same spectrograph
and setup, i. e. to degrade its resolution to the actual resolution 
of the observations. Actually the spectral resolution changes both 
with the position in the field of view and with the wavelength (thus it
is not a mere operation of convolving with the LSF). Taking into account
these effects is particularly critical when, as it is the case here,
the physical velocity dispersion is of the same order or smaller
than the instrumental velocity dispersion.

The procedure for properly taking into account the LSF goes in two steps.
First, determine the LSF as a function of the position in the field
and of the wavelength. Second, inject this LSF in the grid of SSP.

Therefore we made an exhaustive analysis of the LSF of our observations. 
A previous study of the change of the resolution of the MPFS over
the field of view (Moiseev 2001) qualitatively agrees with our results.

To measure the LSF change over the field of view we used the spectra
of standard stars (HD~135722 and HD~175743)
and twilight sky that we analysed with our fitting procedure.
The  high-resolution spectra ($\Delta \lambda = 0.55$\AA; $R \approx 10000$)
for the corresponding stars (the Sun for the twilight spectra) taken from the 
ELODIE.3 library (Prugniel \& Soubiran 2001, 2004), were used as templates.
Since these spectra have exactly the same resolution as the PEGASE.HR
SSPs, the 'relative' LSF that we determined in this way can be directly
injected to the grid of SSP to make it consistent with the MPFS observations.
We parametrize the LSF using $v,
\sigma, h3$ and $h4$.

The whole wavelength range (4100\AA -- 5650\AA) was splitted into five
parts, overlapping by 10 per cent, and the LSF parameters were extracted in each
part independently in order to derive the wavelength dependence of the LSF.

Finally, to inject the LSF in the grid of SSPs, we applied the following
steps to every fiber: (i) Five convolved SSP grids were created using the
 LSF measured for the five wavelength ranges. (ii) The final grid was generated
 by  linear interpolation at each wavelength point between the five grids of
 SSPs. For each fiber of the spectrograph this produces one grid of models
 matched to the LSF of the observations.

\subsection{Results: kinematics, age and metallicity maps}
\label{subsecresssp}
We applied Voronoi adaptive binning procedure to our data, 
setting the target signal-to-noise ratio to 15. The resulting
tessellation includes 76 bins with sizes from 1 to 12 pupils. To get a better
presentation one may interpolate the computed values of each parameter over the
whole field of view using the intensity-weighted centroids of the bins as the
nodes.

We measured the systemic radial velocity $588\pm5$km~s$^{-1}$. The
uncertainty includes possible systematic effects not exceeding 4~km~s$^{-1}$.

\begin{figure*}
\hfil
\begin{tabular}{c c c}
 \includegraphics[width=5.5cm]{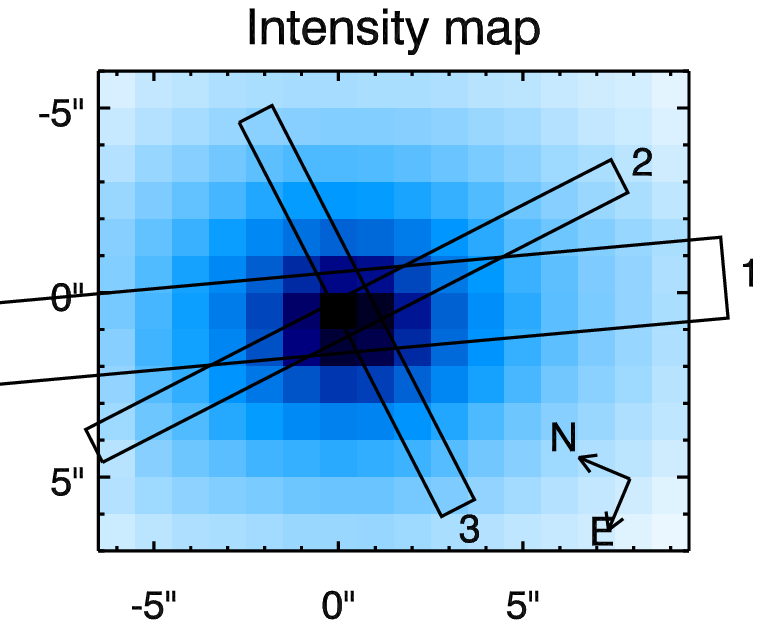} &
 \includegraphics[width=5.5cm]{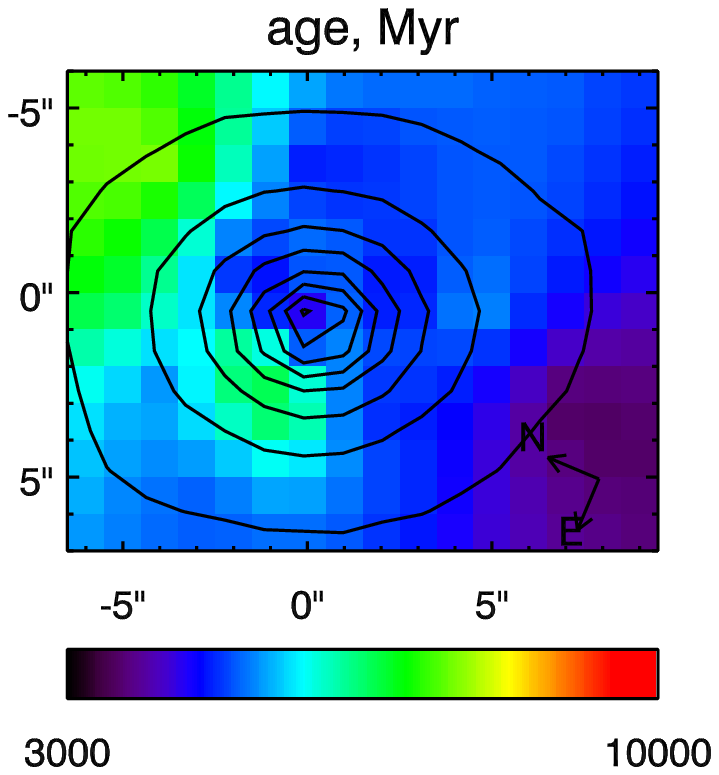} &
 \includegraphics[width=5.5cm]{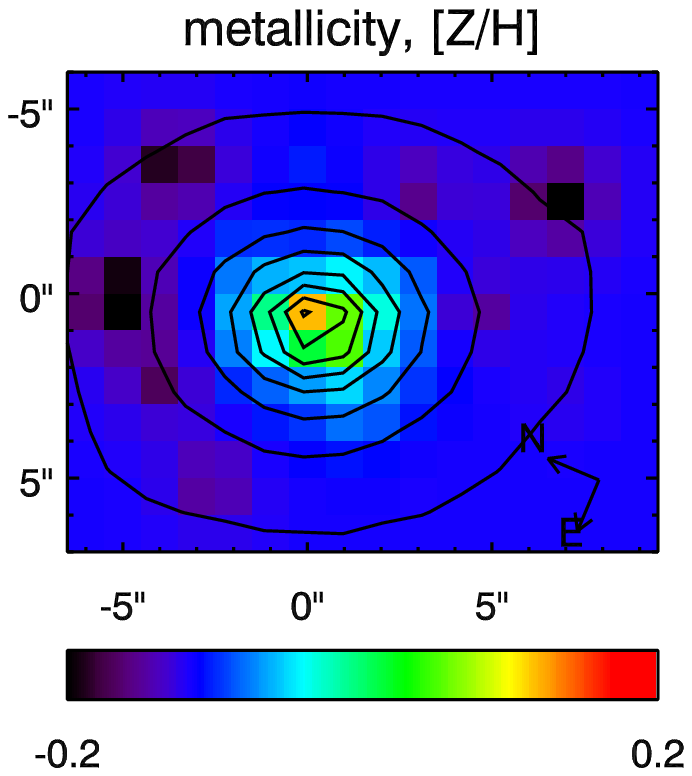} \\
 (a) & (b) & (c) \\
 \includegraphics[width=5.5cm]{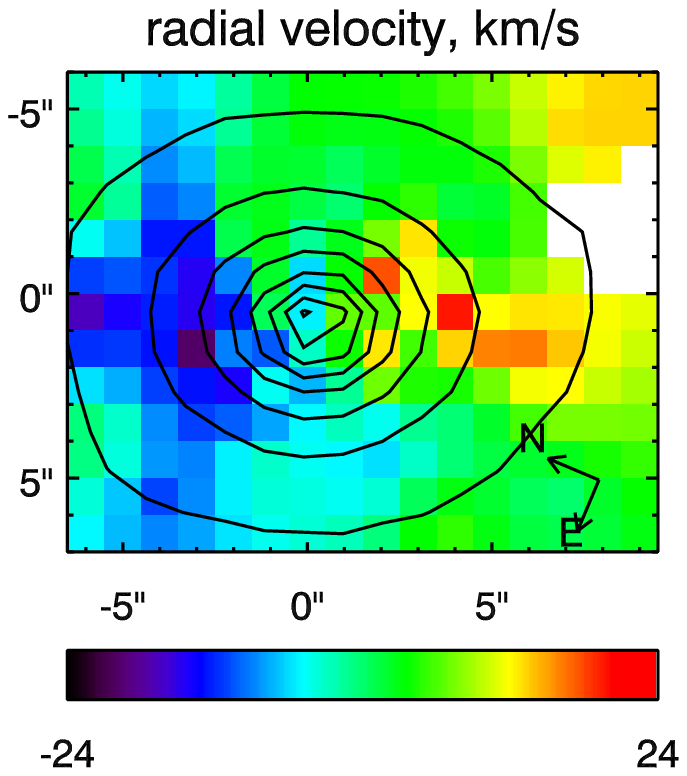} &
 \includegraphics[width=5.5cm]{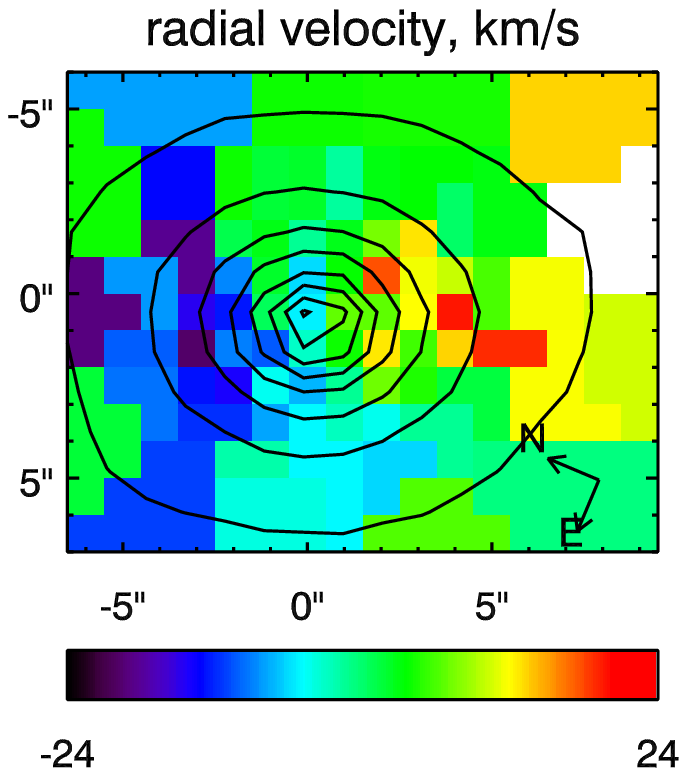} &
 \includegraphics[width=5.5cm]{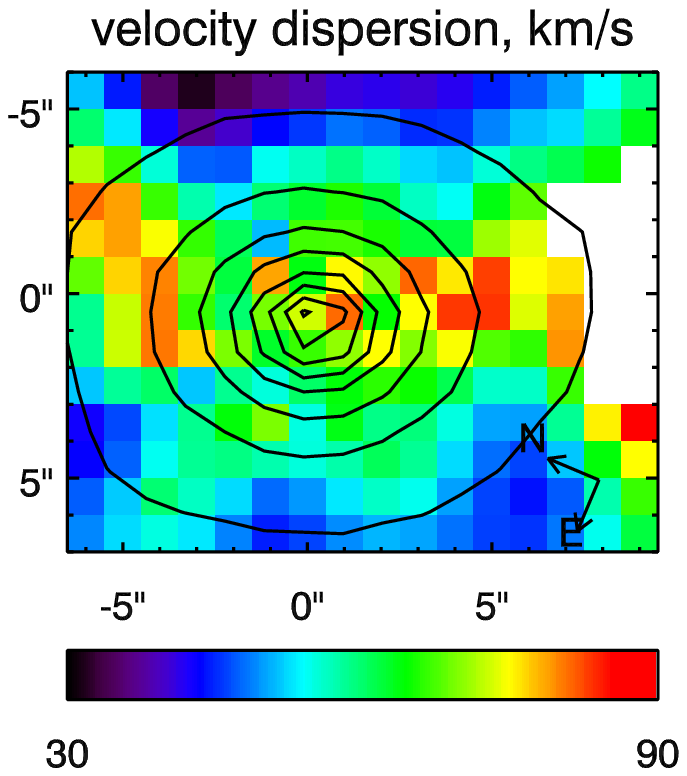} \\
 (d) & (e) & (f) \\
\end{tabular}
 \caption{The generalized view of the kinematical and stellar population data.
(a) intensity map with the positions the kinematical profiles are presented
for: (1) position of slit in Simien \& Prugniel (2002), (2) and (3) positions
of major and minor axis of the embedded stellar disc; 
(b) map of the luminosity-weighted age distribution (in Myr);
(c) map of the luminosity-weighted metallicity 
distribution ($[Z/\mbox{H}]$, dex);
(d) radial velocity field, interpolated between the nodes of the Voronoi
tessellation; (e) radial velocity field: values exactly correspond to the
Voronoi tessellae; (f) velocity dispersion field, interpolated.
\label{fields}
}
\hfil
\end{figure*}

The maps of radial velocity and velocity dispersions are presented in
Fig.~\ref{fields}(d,e,f). The galaxy shows significant rotation and highly inclined disc-like
structure. The uncertainties of the velocity measurements were estimated using
Monte Carlo simulations and confirmed by 
$\chi^2$ maps obtained fitting only multiplicative polynomial continuum
on a grid of values of age, metallicity and velocity dispersion (see
appendix for details).
They depend on the signal-to-noise ratio and change from 2.5~km~s$^{-1}$ for the
$S/N=30$ to 8~km~s$^{-1}$ for the $S/N=10$.

The velocity dispersion distribution shows a gradient from 45-50~km~s$^{-1}$
near the maxima of rotation to 75~km~s$^{-1}$ in the core. Note a sharp peak of
the velocity dispersion of 88~km~s$^{-1}$ slightly shifted to the south-west of the
photometric core. The uncertainties of the velocity dispersion measurements are
3.8~km~s$^{-1}$ for the $S/N=30$, 5.5~km~s$^{-1}$ for the $S/N=20$, and 11~km~s$^{-1}$ for the
$S/N=10$.

\begin{figure}
 \includegraphics[width=8.3cm]{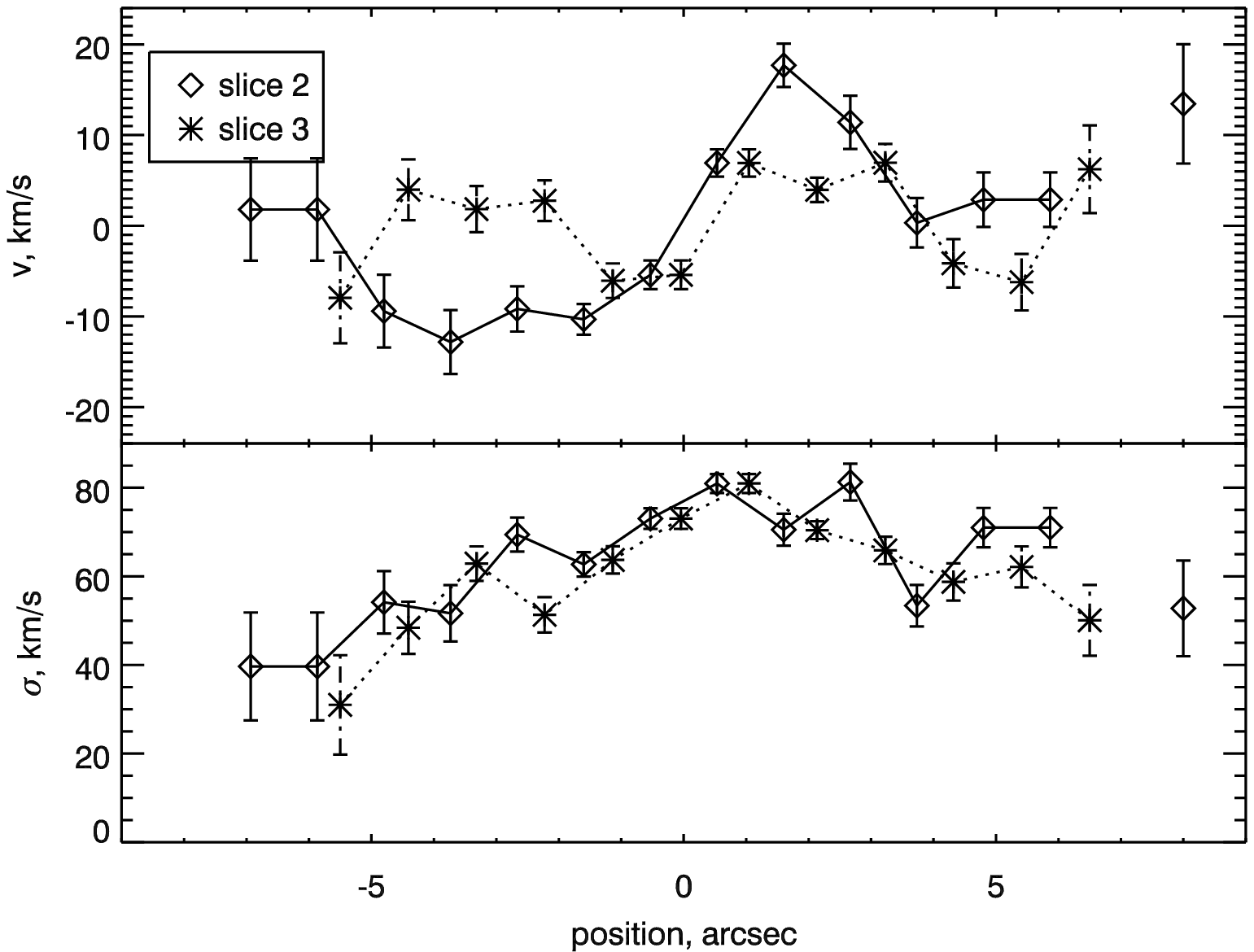}
 \includegraphics[width=8.3cm]{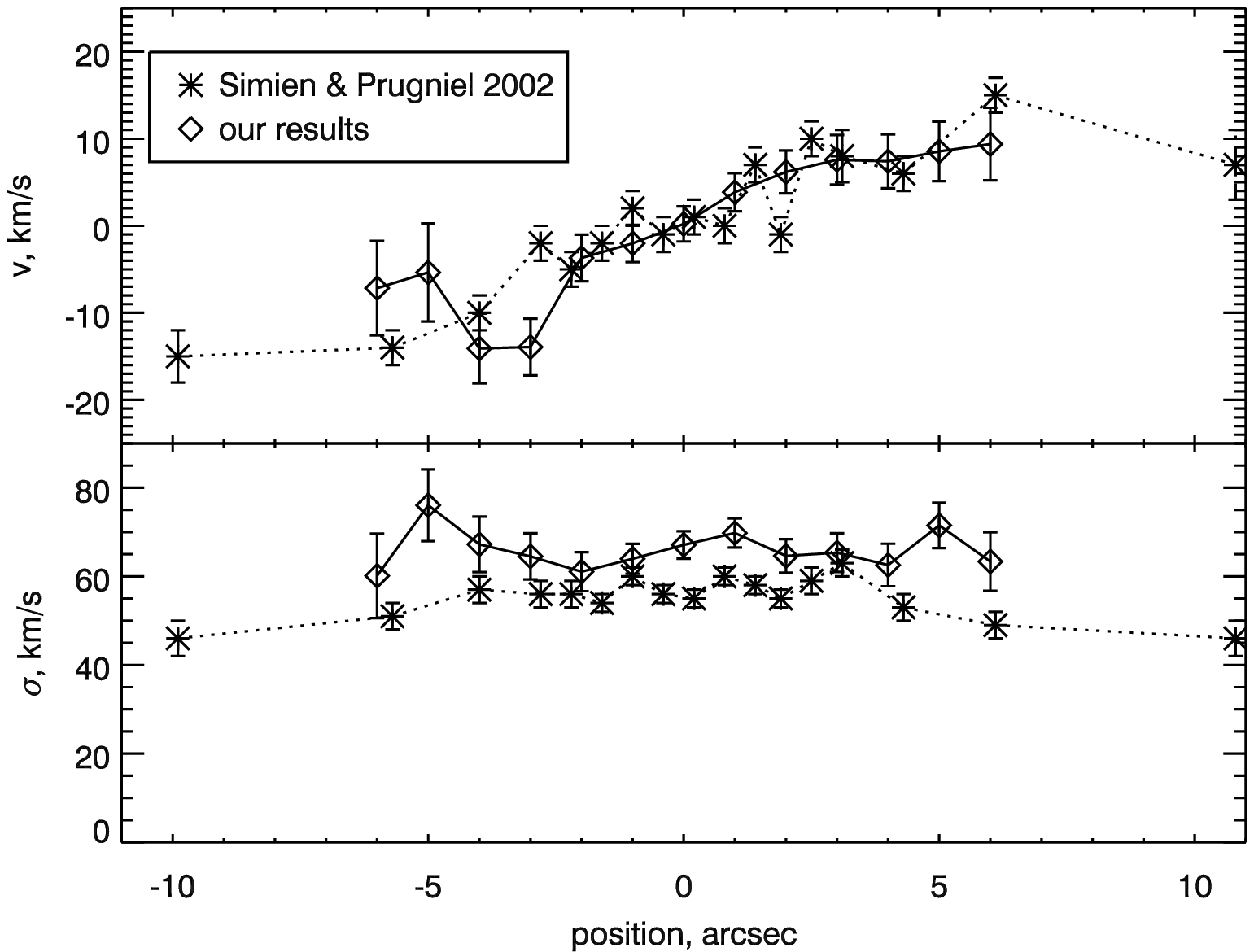}
 \caption{Top pair of plots:
The radial velocity profile for the slice "2"\ (major axis of the embedded
disc) and velocity dispersion profiles for the slices "2"\ and "3" (major
and minor axes of the embedded disc).  Bottom pair of plots: comparison of
the kinematical profiles (slice "1", major axis of main galactic body) to
Simien \& Prugniel (2002): radial velocity and velocity dispersion.
\label{prof}
}
\end{figure}

\begin{figure*}
\hfil
\begin{tabular}{c c}
 \includegraphics[width=4cm]{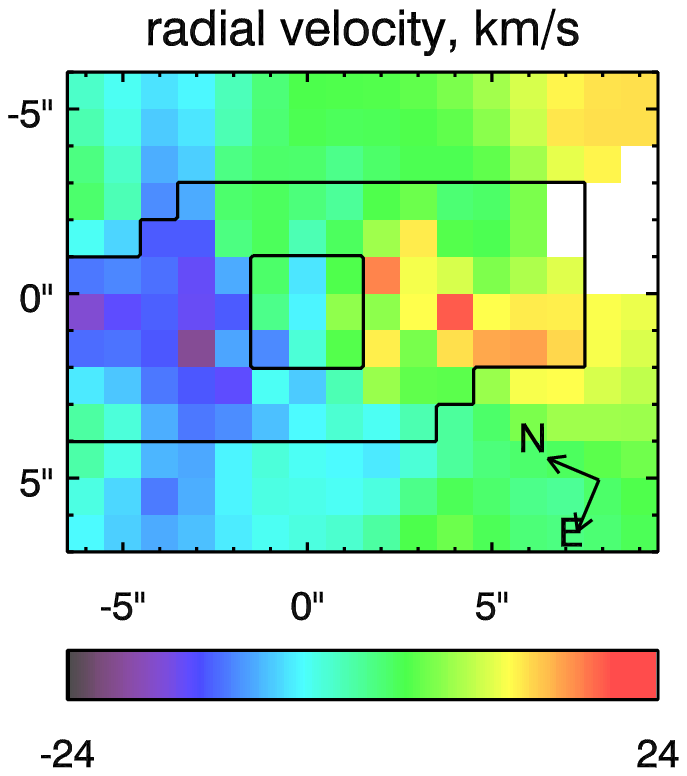} &
 \includegraphics[width=13cm]{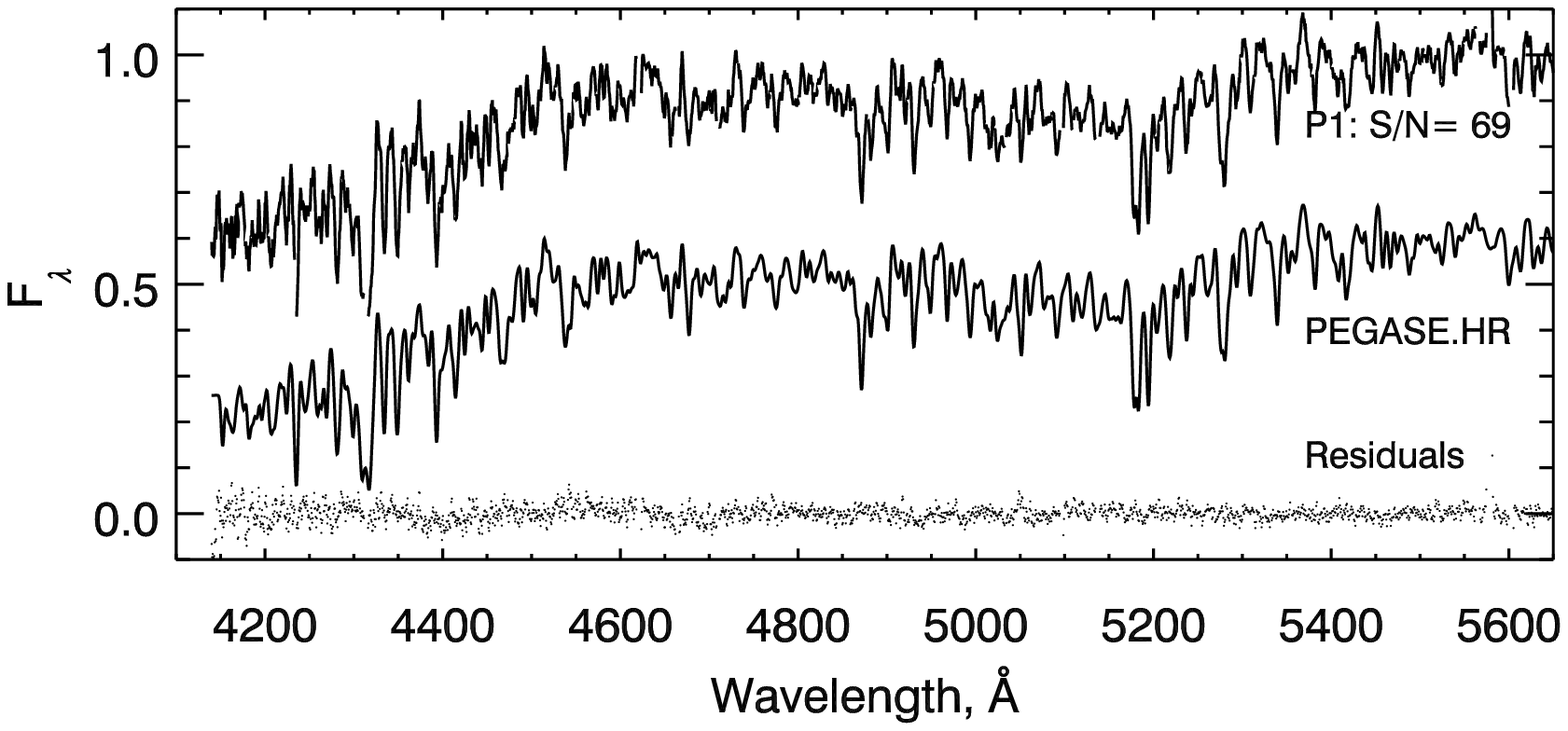} \\
 (a)&(b)\\
\end{tabular}
\begin{tabular}{c}
 (c)\\
 \includegraphics[width=18cm]{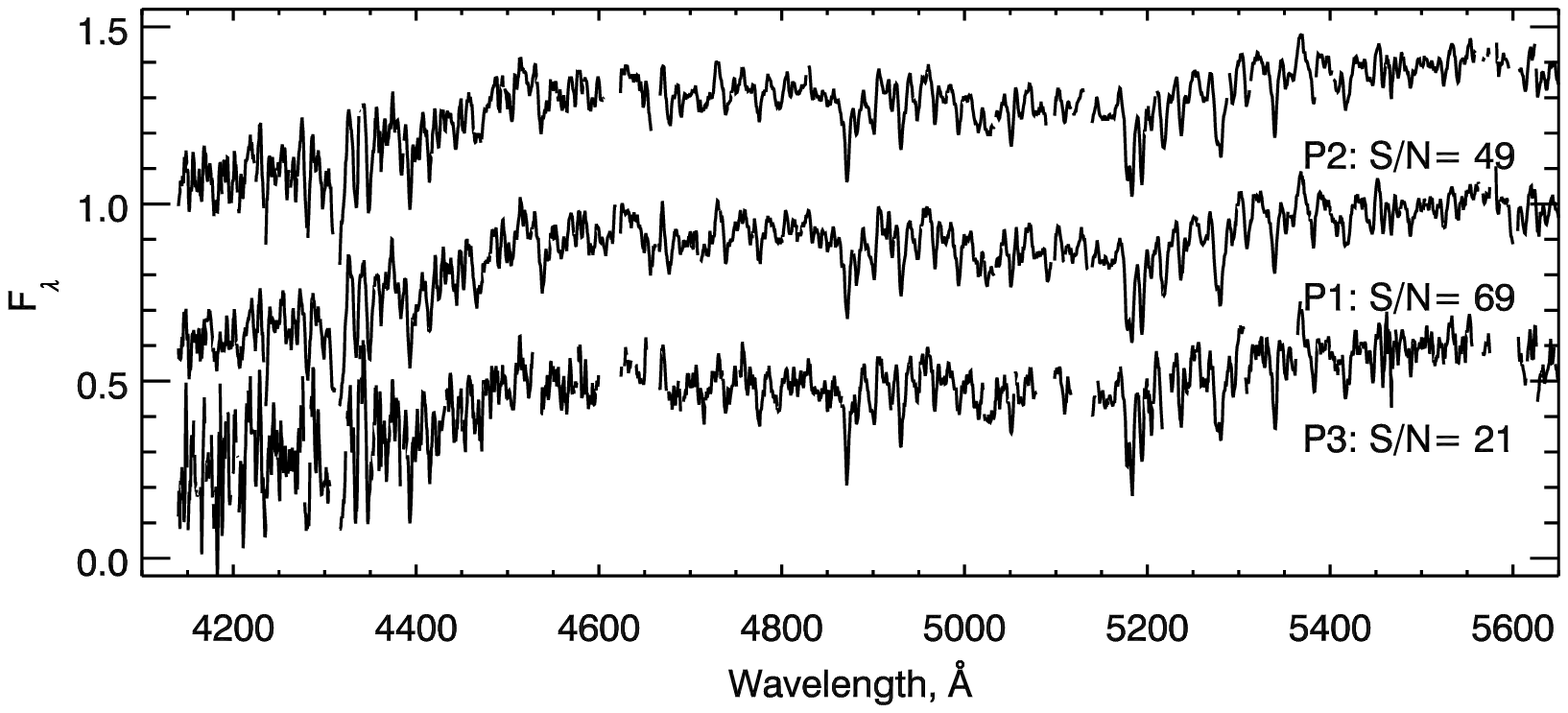}\\
\end{tabular}
 \caption{(a)The radial velocity field with 3 bins overplotted (3-points binning,
 see text); (b) fit of the "P1"\ bin, representing the spectrum, "PEGASE.HR"\
 marking the template spectrum (shifted by -0.3 on the Flux axis from its real
 position, and "Residuals"\ showing the difference between the fit and observed
 spectrum; (c) "P1", "P2", and "P3": co-added spectra for the 3-points binning.
\label{spec4}
}
\hfil
\end{figure*}

The rotation velocity and velocity dispersion profiles are shown in
Fig.~\ref{prof} (top pair). The rotation curve is raising to 20~km~s$^{-1}$ at
the edge of the field (i. e. at a radius of 5~arcsec).

The previous study of IC~3653 was made using long-slit
spectroscopy (Simien \& Prugniel, 2002) under poor atmosphere conditions (6
arcsec seeing). After the proper degrade of the spatial resolution of the MPFS
data the agreement with these earlier observations, both for radial velocity and
velocity dispersion profiles, is excellent (Fig.~\ref{prof}, bottom pair).

\section{Photometry and morphology from ACS images}
\label{secphot}

We have used ACS images from the HST archive, proposal 9401,
"The ACS Virgo Cluster Survey"\ by Patrick C\^ot\'e. The photometric analysis of
IC~3653 is not included in C\^ot\'e et al. 2004. We have converted ACS counts
into corresponding ST magnitudes according to the ACS Data Handbook, available
on-line on the web-site of STScI, and calibrated into AB magnitudes. The derived
colours agree perfectly with those given in Ferrarese et al. (2006).

The F475W--F850LP colour map (equivalent to SDSS $g$-$z$, Fig.~\ref{figcolmaphst})
reveals an elongated region with a major axis of about 7~arcsec, axis ratio $\sim3.5$,
and orientation coinciding with the kinematical disc-like feature, which is
redder than the rest of the map. This colour map was obtained using Voronoi 2D
binning technique applied to the F850LP image in order to reach the
signal-to-noise ratio of 80 per bin.

In the age and metallicity maps derived from the MPFS data, this elongated
feature is seen as a metallicity gradient because of the lower spatial
resolution due to the adaptive binning. The F475W--F850LP colour difference
between the nucleus and the surrounding region 5 arcsec away from the center is
about 0.05 mag. Using pegase.2 templates (Fioc \& Rocca-Volmerange, 1997
\footnote{http://www2.iap.fr/users/fioc/PEGASE.html}) and ACS filters
transmission curves (Sirianni et al. 2005) we find that this colour difference
corresponds to a metallicity difference of about 0.1 dex (assuming an age of 5
Gyr as found from our spectroscopic analysis), a value consistent with the 
spectroscopic analysis.

We have fitted two-dimensional S\'ersic profile using the GALFIT package
(Peng et al. 2002). We can see significant positive residuals, representing the
nucleus in the very centre (around 1.5 arcsec in size, central surface
brightness ST$_{475} \sim 16.25$~mag~arcsec$^2$, slightly asymmetric and
offcentered with respect to the centre of the S\'ersic profile having n=1.88,
$R_e$=6.9~arcsec, and $\epsilon$=0.11 (S\'ersic index, effective radius, and
ellipticity respectively; our values coincide with ones from Ferrarese et al.
2006). There are faint large-scale residuals as well, that can be explained by
superposition of several components (at least two).

\begin{figure}
\includegraphics[width=8.3cm]{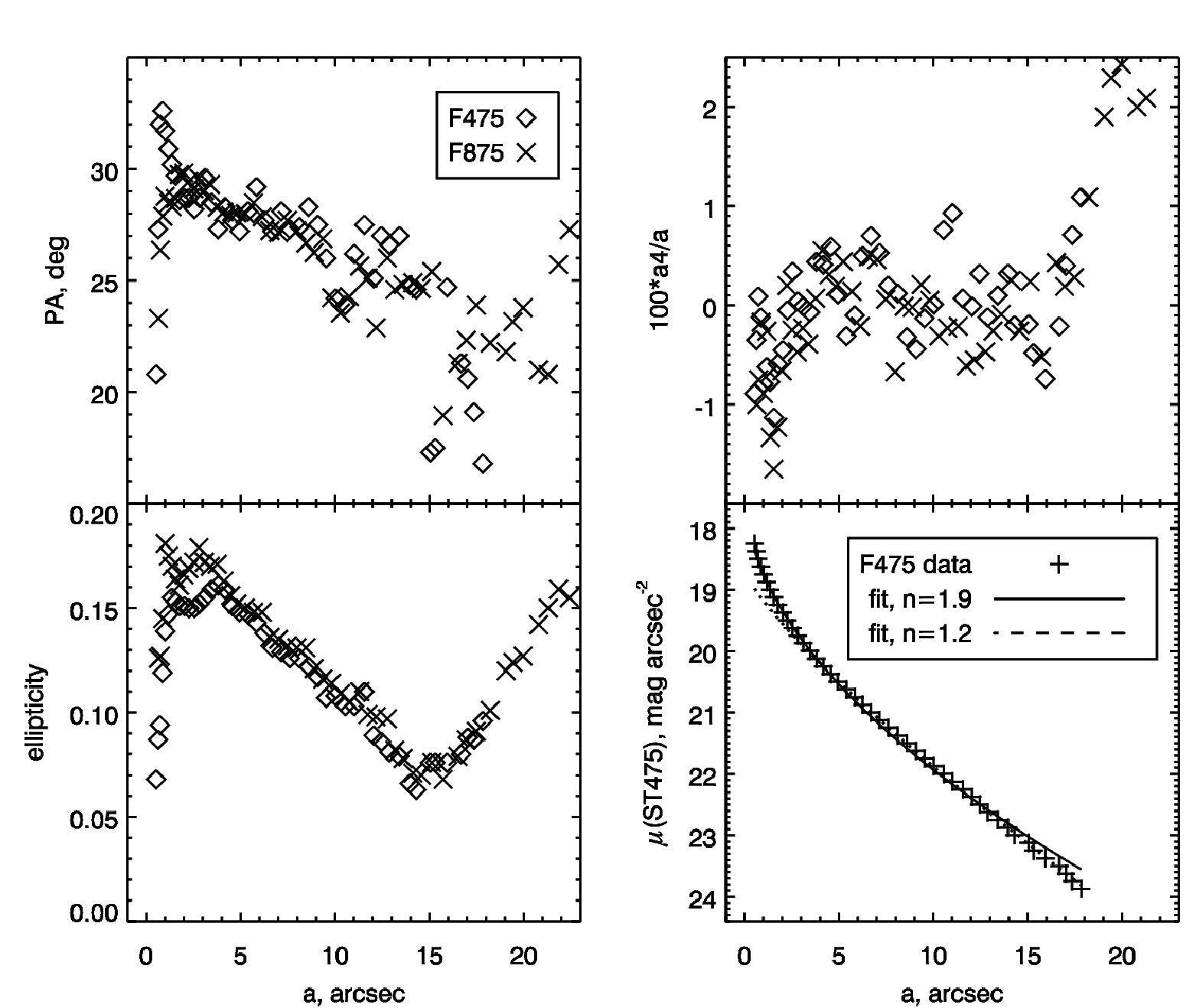}
\caption{Photometrical characteristics of IC~3653: position angle (upper
left), ellipticity of the isophotes (lower left), disky/boxy parameter a4 (upper
right) in two colours, and photometric profile in F475W. 
\label{figphotpar}
}
\end{figure}

\begin{figure}
\includegraphics[width=8.3cm]{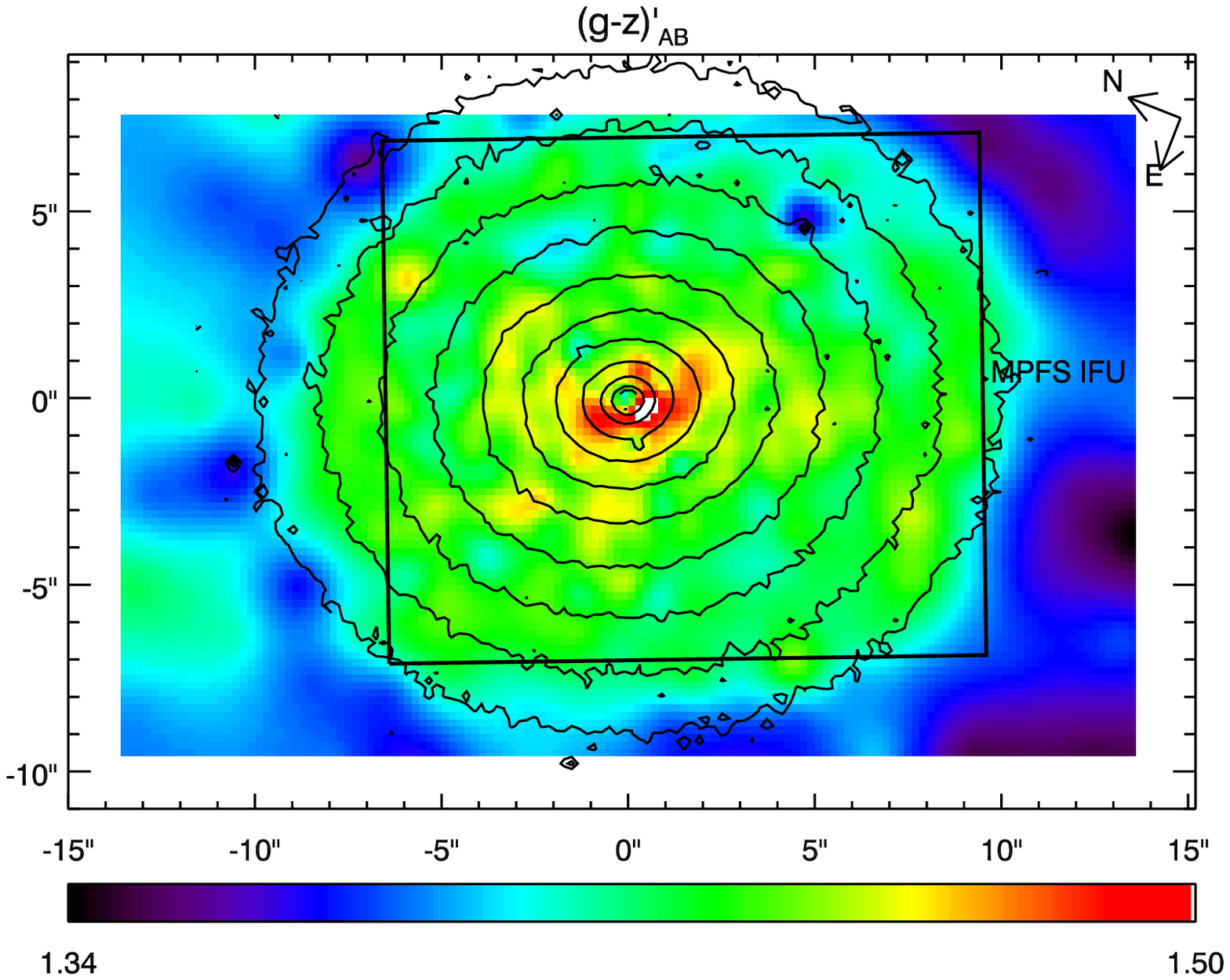}
\caption{F475W-F850LP colour map (HST ACS data). Isophotes of the F850LP image are
overimposed. The position and size of the MPFS field of view are shown.
\label{figcolmaphst}
}
\end{figure}

Then we have modeled the images by elliptical isophotes with free center and
orientation. We see some isophote twist and ellipticity change in the
inner region of the galaxy that may be due to the embedded disc. Main parameters of
the model fitted are presented in Fig.~\ref{figphotpar}. The subtraction of the
model from the original image does not reveal any internal feature.

On the lower right plot in Fig.~\ref{figphotpar} the F475W light profile
is shown with crosses. The solid line represents the best-fitting S\'ersic
profile for the whole galaxy excluding only very centre (inner 1 arcsec) with
n=1.9, and dashed line gives the best-fitting (n=1.2) for the periphery of
the galaxy outside 7~arcsec, i. e. beyond the disc found in the colour map. We
consider this latter fit as the most representative.

\section{Discussion}
\label{secdisc}
Both line-of-sight velocity field extracted from the MPFS data cube and
colour map obtained from the HST imagery provide coinciding
arguments for a presence of a faint internal co-rotating stellar disc
embedded within a rotating spheroid. This is the main observational result
from our study of IC~3653, which may be regarded to some extend as an edge-on
counterpart of IC~3328, the dE with embedded spiral structure found by Jerjen et
al. (2000). The disc of IC~3328 is detected to a larger radius than the one of
IC~3653: the spiral structure is seen up to $\approx$ 30~arcsec = 2~$r_e$, while
the disc of IC~3653 is detected only to 3~arcsec = 0.5~$r_e$. The contrast of
the spiral in IC~3328 is a few per cent and in IC~3653 the disc is seen from the
colour and only as a weak maximum of ellipticity at the radius of its maximum
contrast to the bulge. In both galaxies, the discs are essentially stellar (no
sign of star formation) and old. Estimating and comparing the mass of the two
discs is strongly model-dependent, and beyond the scope of this study.

In this section we will first compare the results of two methods for estimating
stellar population parameters: Lick indices and pixel-fitting, then the
characteristics of IC~3653 to other dE galaxies. Finally, we will review the
different possible origins of the particular properties of this galaxy.

\subsection{Comparison of two techniques for estimating stellar population
characteristics}
In Table~\ref{tabcomptz3b} we present comparison between SSP-equivalent age
and metallicity obtained with the pixel fitting and inversion
of bi-index grids for H$\beta$-Mg$b$, H$\beta$-$<$Fe$>'$, H$\beta$-[MgFe]
using index measurements on observed spectra and best-fitting templates.

One can notice particularly good agreement between the approaches. The ages
derived from Lick indices appear to be slightly older, but the difference is
not significant. The best agreement for both ages and metallicities is
reached between pixel fitting and measurements of H$\beta$ and the combined
[MgFe] index (Thomas et al. 2003). The internal precision of the parameters
derived from pixel fitting is better those from Lick indices by a factor
three to four, depending on the indices used.  This can be explained by the 
optimized usage of the information by the pixel fitting procedure. Though it is
difficult to assess the reliability of these small error bars, the relative
variations of age and metallicity can be trusted, even when the signal-to-noise
ratio is as low as 10 per pixel (with MPFS spectral resolution and wavelength
coverage).

\begin{table}
\begin{tabular}{l c c c}
 & "P1" & "P2" & "P3" \\
\hline
\hline
t$_{fit}$, Gyr &  4.93$\pm$ 0.20 &  4.95$\pm$ 0.30 &  4.97$\pm$ 0.70 \\
\hline
t$_{H\beta-Mgb}$ &  7.04$\pm$ 1.56 &  6.97$\pm$ 1.47 & 11.25$\pm$ 6.02 \\
t$_{H\beta-<Fe>'}$ &  7.02$\pm$ 1.33 &  7.28$\pm$ 2.20 &  6.08$\pm$ 3.94 \\
t$_{H\beta-\mbox{[MgFe]}}$ &  7.11$\pm$ 1.65 &  6.88$\pm$ 1.89 &  6.70$\pm$ 5.11 \\
\hline
t$_{H\beta-Mgb\mbox{mod}}$ &  5.27$\pm$ 1.56 &  5.13$\pm$ 1.47 &  4.85$\pm$ 6.02 \\
t$_{H\beta-<Fe>'\mbox{mod}}$ &  5.15$\pm$ 1.33 &  4.30$\pm$ 2.20 &  4.23$\pm$ 3.94 \\
t$_{H\beta-\mbox{[MgFe]mod}}$ &  5.22$\pm$ 1.65 &  4.12$\pm$ 1.89 &  4.15$\pm$ 5.11 \\
\hline
\hline
Z$_{fit}$, dex &  0.03$\pm$ 0.01 & -0.14$\pm$ 0.02 & -0.17$\pm$ 0.05 \\
\hline
Z$_{H\beta-Mgb}$ & -0.05$\pm$ 0.09 & -0.18$\pm$ 0.08 & -0.34$\pm$ 0.24 \\
Z$_{H\beta-<Fe>'}$ & -0.02$\pm$ 0.04 & -0.10$\pm$ 0.05 &  0.03$\pm$ 0.13 \\
Z$_{H\beta-\mbox{[MgFe]}}$ & -0.04$\pm$ 0.05 & -0.14$\pm$ 0.06 & -0.11$\pm$ 0.15 \\
\hline
Z$_{H\beta-Mgb\mbox{mod}}$ & -0.02$\pm$ 0.09 & -0.14$\pm$ 0.08 & -0.16$\pm$ 0.24 \\
Z$_{H\beta-<Fe>'\mbox{mod}}$ &  0.02$\pm$ 0.04 & -0.14$\pm$ 0.05 & -0.17$\pm$ 0.13 \\
Z$_{H\beta-\mbox{[MgFe]mod}}$ &  0.00$\pm$ 0.05 & -0.13$\pm$ 0.06 & -0.16$\pm$ 0.15 \\
\hline
\hline
\end{tabular}
\caption{Comparison of age and metallicity measurements for the 3-points binning
obtained with pixel fitting and based on different pairs of Lick indices grids: 
measured on real spectra and on best-fitting templates.
\label{tabcomptz3b}
}
\end{table}

\subsection{Properties and nature of IC~3653}

We computed the position of IC~3653 on the fundamental plane (FP, Djorgovski \&
Davis 1987). "Vertical"\ deviation from FP ($d=-8.666 + 0.314 \mu_e + 1.14 \log
\sigma_0 - \log R_e$, Guzman et al. 1993) is $d=0.2$. This deviation places
IC~3653 into the centre of the cloud, representing dE galaxies in De Rijcke et
al. (2005, their figure 2, left panel), and exactly on the theoretical
predictions by Chiosi \& Carraro (2002) and Yoshii \& Arimoto (1987),
overplotted on the same figure.

The mean age of the stellar population of IC~3653, $t=5$~Gyr, coincides with the
mean age of dE galaxies in Virgo: $t_{mean}=5$~Gyr in Geha et al. (2003);
$t_{mean}=5...7$~Gyr in Van Zee et al. (2004b). By contrast, the metallicity of
the main body, $Z=-0.1$ is slightly higher than in these other galaxies: 
$Z_{mean}=-0.3$ in Geha et al. (2003); $Z_{mean}=-0.4$, Van Zee et al. (2004b).
It may be understood as a consequence of the luminosity-metallicity relation if
we remember that IC~3653 is more luminous than most of the galaxies in the
samples of Geha et al. (2003), and Van Zee et al. (2004b).

We see that fundamental properties of IC~3653 do not differ from typical dE
galaxies, though the effective radius is one of the smallest within the samples
of Virgo dE's presented in Simien \& Prugniel (2002), Geha et al. (2003), and
van Zee (2004a).

We derived the B-band mass-to-light ratio of IC~3653 following the method by
Richstone \& Tremaine (1986) as $M/L_B = 8.0\pm1.5 (M/L_B)_{\odot}$. The stellar
mass-to-light ratio of the best fitting PEGASE.HR model in the "P1"\ region is
$(M/L_B)_{*}=3.7\pm0.4 (M/L_B)_{\odot}$. If we trust this simple dynamical
model, it indicates that the dark matter content of IC~3653 is similar to
recently found in De Rijcke et al. (2006) for the M~31 dE companions, i. e.
about 50 per cent.

Recent theoretical studies based on N-body modelling of the evolution of disc
galaxies within a $\Lambda$ CDM cluster by Mastropietro et al. (2005)
suggest that discs are never completely destroyed in cluster environment, even
when the morphological transformation is quite significant. Our discovery of the
faint stellar disc in IC~3653 comforts these results. Thus, one of the possible
origins of IC~3653 is morphological transformation in the dense cluster
environment from late-type disc galaxy (dIrr progenitor). Gas was removed by
means of ram pressure stripping and star formation was stopped. This process
must have finished at least 5~Gyr ago, otherwise we would have seen a younger
population in the galaxy. However, duration of the star formation period must
have been longer than 1~Gyr, otherwise deficiency or iron (i.e. Mg/Fe
overabundance) would have been observed. Within this scenario, metallicity
excess in the disc can be explained by a slightly longer duration of the star
formation episode compared to the spheroid. But we cannot see the difference
in the star formation histories (even luminosity weighted age), because of
insufficient resolution on the stellar population ages.

Another way to acquire a disc having higher metallicity than the rest of
the galaxy is to experience a minor dissipative merger event (De Rijcke et
al. 2004).  
This is
rather improbable for a dwarf galaxy, but cannot be completely excluded. In
particular, kinematically decoupled cores, recently discovered in dwarf
and low-luminosity
galaxies (De Rijcke et al. 2004, Geha et al. 2005, Prugniel et al. 2005,
Thomas et al. 2006) can be explained by minor merger events.

The most popular scenario which is usually considered to explain
formation of embedded stellar discs in giant early-type galaxies is star
formation in situ after infall of cold gas onto existing rotating spheroid,
e.g. from a gas-rich companion (cross-fueling). This scenario has been
used by Geha et al. (2005) to explain counter-rotating core in NGC~770, a
dwarf S0 which is more luminous than IC~3653 ($M_B=-18.2$~mag) and located
close to a massive spiral companion NGC~772 ($M_B=-21.6$~mag) in a group. In
NGC~127 ($M_B=-18.0$~mag), a close satellite of another giant gas-rich galaxy,
NGC~128, we observe the process of cross-fueling at present (Chilingarian,
2006). Group environment, where relative velocities of galaxies are rather low,
favours of the interaction processes on large timescales, such as smooth gas
accretion.

Our data on IC~3653 cannot provide a decisive choice between those
alternatives. However, from a general point of view, dynamically hot
environment of the Virgo does not favour the slow accretion of cold gas.
IC~3653 is not a member of a subgroup including large galaxies, which can foster
gaseous disc, so we believe that in this particular case the scenario of slow
accretion is not applicable.

At present, the sample of objects where disc-like sub-structures were
searched either from images or integral field spectroscopy is still too
small to draw statistical conclusions. But it is quite probable that the
progenitors of the dE's were disc galaxies (pre-dIrr or small spiral galaxies)
and that they evolved due to feedback of the star formation and
environmental effects. Present dIrr also experienced feedback but kept
their gas, so it is unlikely that the feedback alone can remove the
gas. Therefore environmental effects are probably the driver of the
evolution of dE's, and the discovery of stellar discs in dE's is consistent
with this hypothesis.

\section*{Acknowledgments}
\label{secack}

We are very grateful to Alexei Moiseev for supporting the observations
of IC~3653 at the 6-m telescope.
Visits from PP in Russia and IC in France were supported through a CNRS
grant. PhD of IC is supported by the INTAS Young Scientist Fellowship
(04-83-3618). The dwarf galaxies investigation by IC, OS, and VA
is supported by the bilateral Flemish-Russian collaboration (project 
RFBR-05-02-19805-MF\_a). We also appreciate support by IAU for
attending the IAU Colloquium 198, where preliminary results of this paper
were presented. We would like to thank our colleagues: Francois Simien and
Sven de Rijcke, and students: Mina
Koleva and Nicolas Bavouzet for fruitful discussions. Special thanks to the
Large Telescopes Time Allocation Committee or the Russian Academy of Sciences
for providing observing time with MPFS.

\appendix

\section{Validation and error analysis}
\label{secvalid}

In this appendix we address questions concerning the error analysis, stability
of the solutions, and possible biases for relatively old stellar population
(about 5~Gyr). A full description of these aspects extended to any instrument
and much wider range of parameters will be described in details in the 
forthcoming paper (Koleva et al. in preparation). Here we give only essential
error analysis required for validation of the results presented in this paper
and in the forthcoming papers based on MPFS data for dwarf galaxies.

\subsection{Error analysis}
\label{subsecerr}
Error estimates on the stellar population parameters is complicated because of
the internal degeneracies and coupling, and the questions of unicity of the
solution and local minima are always critical. A complete and detailed
description of one of the possible approaches to locate the alternate solutions
using for a allied inversion technique is given in Moultaka \& Pelat (2000).

We performed some Monte-Carlo simulations (about 10000 per spectrum for the
3-points binning) to demonstrate the consistency between the uncertainties on
the parameters reported by the minimization procedure and the real error 
distributions. These simulations have demonstrated that in case of our dataset,
where there is neither significant model mismatch due to element abundance
ratios, nor strong mistake with subtraction of additive terms (diffuse light and
night sky), the uncertainties found from the Monte-Carlo simulations using
scattering of solutions in the multidimensional parameter space coincide with
values reported by the minimization procedure being multiplied by $\chi^2$
values. Deviation of $\chi^2$ from 1 might be caused either by poor quality
of the fit (model-mismatch) or by wrong estimations of absolute flux 
uncertainties in the input data. We inspected the residuals of the fit visually
and by averaging them using wide smoothing window (150 pixels) and found no 
significant deviations between model and observations. Thus we conclude that in
some cases our estimations of the uncertainty on the absolute fluxes based on
the photon statistics may be wrong by 15 per cent, resulting in values of
$\chi^2$ between 0.7 and 1.3.

To estimate the errors more accurately, locate possible alternate solutions and
search for degeneracies between kinematical and stellar population parameters
we study the distribution of $\chi^2$ in the age -- metallicity -- velocity
dispersion space. The procedure we followed includes the following steps:
\begin{itemize}
  \item We chose a grid of values for age, metallicity, and velocity dispersion,
  that was supposed to cover a reasonable region of the parameter space where we
  could expect to have solutions. In our case the grid was defined as: 2~Gyr $<$
  t $<$ 14~Gyr with a step of 200~Myr, -0.45$<$[Fe/H]$<$0.40 with a step of
  0.01~dex, 30~km~s$^{-1}$ $<$ $\sigma$ $<$ 100~km~s$^{-1}$ with a step of 0.5~km~s$^{-1}$ 
  \item At every node of t-Z grid we ran the pixel fitting procedure in order
  to determine multiplicative polynomial continuum, and to have the best
  fit for a given SSP
  \item Later $\chi^2$ was computed on the grid of values of $\sigma$ by fixing
  all other components of the solution that had been found in the previous step
\end{itemize}

The reason to scan the $\sigma$ dimension with fixed multiplicative polynomial
is to save computer time by reducing the number of free parameters, and it is
justified by the strong decoupling between the multiplicative polynomial
(affecting only low frequencies) and the velocity dispersion (affecting high
frequencies).

In other words using our procedure we compute a slice of the full $\chi^2$ by
the hypersurface defined as a set of minimal $\chi^2$ values for multiplicative
continuum terms and radial velocity values, and then reproject it onto "t-Z"\ and
"t-Z-$\sigma$"\ hyperplanes. The result contains two arrays: 2D age-metallicity
and 3D age-metallicity-velocity dispersion.

The values of line-of-sight radial velocity obtained during the fitting
procedure on the "t-Z"\ grid are constant within the errors, suggesting that
scanning of the $\chi^2$ hyperspace on $v$ is not necessary.

\begin{figure*}
\hfil
 \includegraphics[width=17cm]{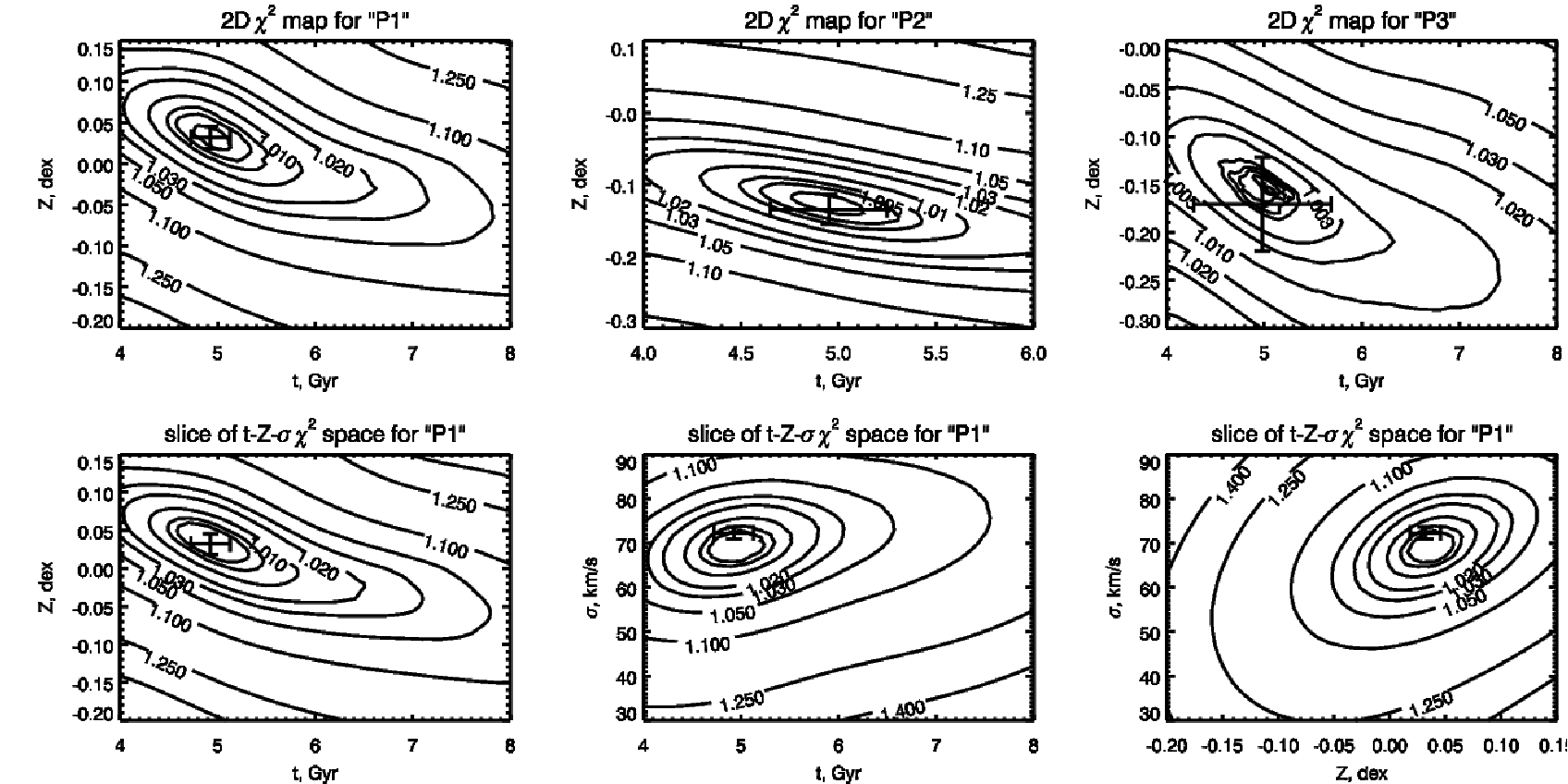}
 \caption{2-dimensional maps of $\chi^2$ distributions for the 3-points binning
 (upper row), and slices of the 3-dimensional $\chi^2$ distribution 
 (t-Z-$\sigma$) for the "P1"\ bin.
\label{figchi2map}
}
\hfil
\end{figure*}

In Fig.~\ref{figchi2map} (upper line) we present the maps of $\chi^2$ for the
3-points binning on the "t-Z"\ plane. One can see elongated shapes of the
minima, corresponding to well known age-metallicity
degeneracy. Three plots on the bottom of Fig.~\ref{figchi2map} represent slices
of the 3D $\chi^2$ space scan (t-Z-$\sigma$) for the "P1"\ bin. One can notice
that the width of the minimum on "t-Z"\ plane has decreased due to a correlation
between metallicity and velocity dispersion, that is clearly seen on the
"Z-$\sigma$"\ slice. This degeneracy between velocity dispersion and metallicity
can be understood as follows: an overestimate of the metallicity in the template
increases depth of the absorption lines and can be compensated by
higher velocity dispersion. In principle an important mismatch of the
metallicity when analysing the kinematics with usual methods may bias the
velocity dispersion and even produce patterns in velocity dispersion profiles,
whereas it is in reality a metallicity gradient. A mismatch of 0.1 dex results
in a bias of 15 per cent on the velocity dispersion.

\subsection{Stability of solutions}
\label{subsecstab}
Stability of solutions is a crucial point for every method dealing with
multiparametric non-linear minimization. We studied the stability with respect to
initial guess, wavelength range being used, and degree of the multiplicative
polynomial continuum.

\subsubsection{Initial guess}
\label{subsubsecstabini}
We have made several dozens of experiments with different initial guesses in
order to inspect the stability of convergence. We found a correct convergence
for a wide range of guesses of age, metallicity and velocity dispersions, but
for radial velocity the initial guess must be within twice the velocity
dispersion from the solution.

\subsubsection{Wavelength range}
\label{subsubsecwlr}
We ran two series of experiments: one with $\lambda > 4700$\AA, and another one
with the full wavelength range, but regions of Balmer lines (H$\gamma$ and 
H$\beta$) masked. The first experiment checks the sensitivity to the wavelength
range (several strong metallic features are excluded) and the second one removes
the best age estimators (Worthey at al. 1994, Vazdekis \& Arimoto 1999), so that
instability on age may be expected.

The first experiment does not reveal any bias, but only an expectable increase
of the uncertainties.

The second experiment does not neither reveal instability, but though the
errors on metallicity and velocity dispersion are not strongly affected, the
error on age is multiplied by 3 (therefore the precision become comparable to
Lick indices). It is remarkable that even without Balmer lines it is possible to
give estimations of age of the stellar population.

\begin{table}
\begin{tabular}{l c c c}
 & P1 & P2 & P3 \\
\hline
$v$,~km~s$^{-1}$ &  601.8$\pm$   1.0 &  603.4$\pm$   1.4 &  603.8$\pm$   3.0 \\
 &  600.9$\pm$   1.0 &  603.1$\pm$   1.7 &  603.7$\pm$   3.4 \\
\hline
$\sigma$,~km~s$^{-1}$ &   70.9$\pm$   1.6 &   67.3$\pm$   2.2 &   52.1$\pm$   5.0 \\
 &   71.8$\pm$   1.6 &   65.3$\pm$   2.6 &   52.1$\pm$   5.7 \\
\hline
t, Gyr &  4.855$\pm$ 0.218 &  4.728$\pm$ 0.289 &  4.629$\pm$ 0.694 \\
 &  4.714$\pm$ 0.235 &  4.448$\pm$ 0.403 &  4.238$\pm$ 0.930 \\
\hline
Z, dex &   0.01$\pm$  0.02 &  -0.14$\pm$  0.02 &  -0.15$\pm$  0.05 \\
 &   0.03$\pm$  0.01 &  -0.13$\pm$  0.02 &  -0.15$\pm$  0.05 \\
\hline
\end{tabular}
\caption{Stability of the solutions for the 3-points
 binning with respect to the wavelength range. First lines for every parameter
 correspond to $\lambda > 4700$\AA, second ones to full range with Balmer lines
 excluded.
\label{stabwlr}
}
\end{table}

\subsubsection{Order of multiplicative polynomial continuum}
\label{subsubsecmult}
\begin{figure}
 \includegraphics[width=8.0cm]{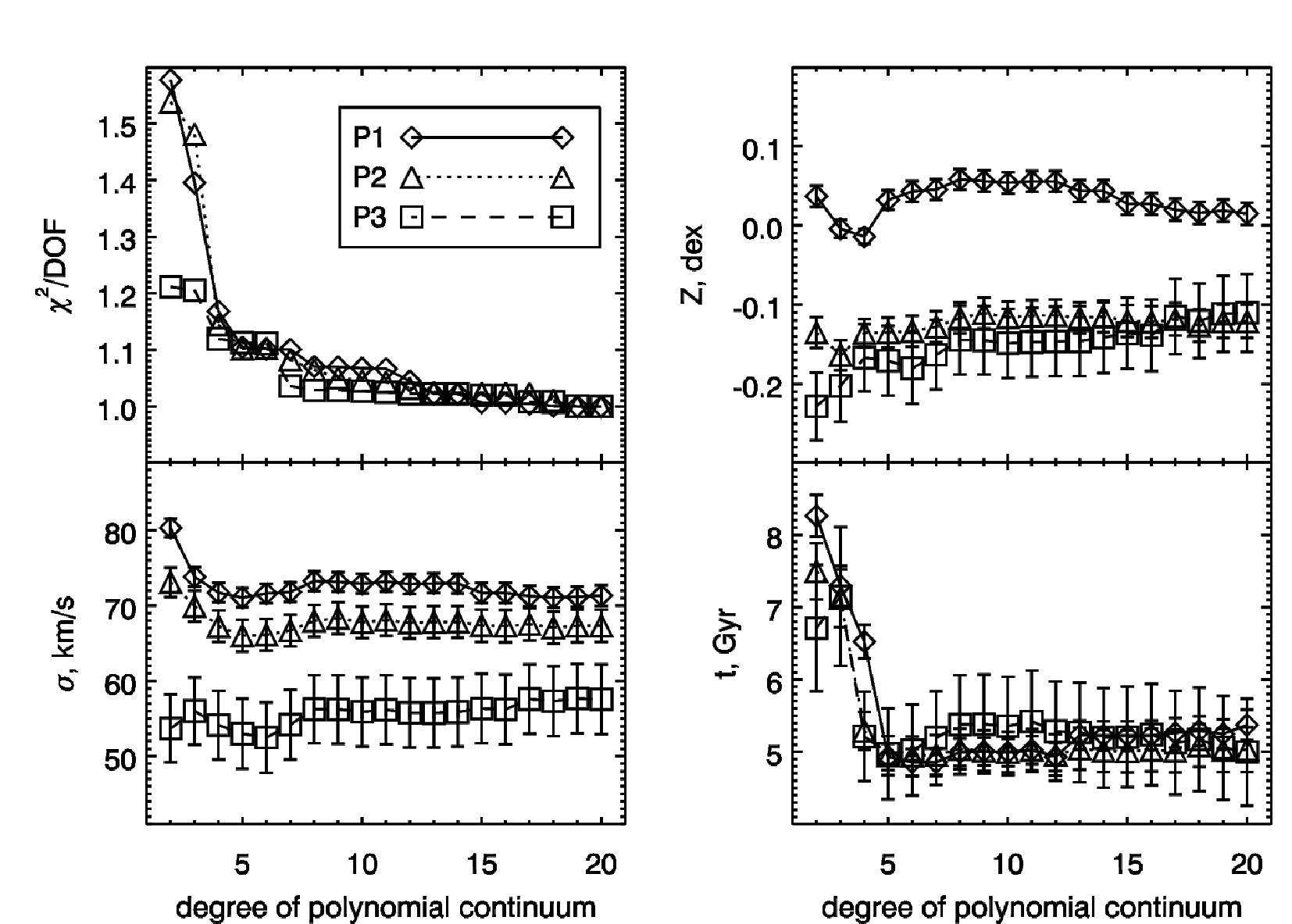}
\caption{Stability of the parameters with respect to the order of the
multiplicative polynomial continuum. Different plotting symbols correspond to
three bins of the 3-points binning.
\label{figstabmult}
}
\end{figure}

We also explored the stability of the method with respect to the order of the
multiplicative polynomial continuum. The results are shown in
Fig.~\ref{figstabmult}. One can see that for n$>$5 there is neither
significant changes of the estimations of kinematical and stellar population
parameters, nor of $\chi^2$ value. Therefore we chose $n=5$ for all our data
analysis.

\subsection{Possible biases}
\label{subsecbias}
There are several possible sources of systematic errors on the parameters: (1)
additive systematics of the flux calibration due to under- or oversubtraction
of the night sky, (2) imperfections of the models, one of the most important
among those is non-solar abundance ratio. Since IC~3653 exhibits exactly solar
[Mg/Fe] (see Fig.~\ref{lickdiag}) everywhere in the field of view, we do not
study possible consequences of abundance ratio mismatch in this paper.

The accurate subtraction of the night sky emission is a challenging step
of the data reduction for low-surface brightness objects. Basically, night sky
emission consists of continuum emission, that might include scattered solar light
as well, and several bright emission lines. Under- or oversubtraction of night
sky brings additive component resulting in changing the relative depths of
spectral features, and therefore affects particularly the estimates of the
metallicity.

\begin{figure}
 \includegraphics[width=8.0cm]{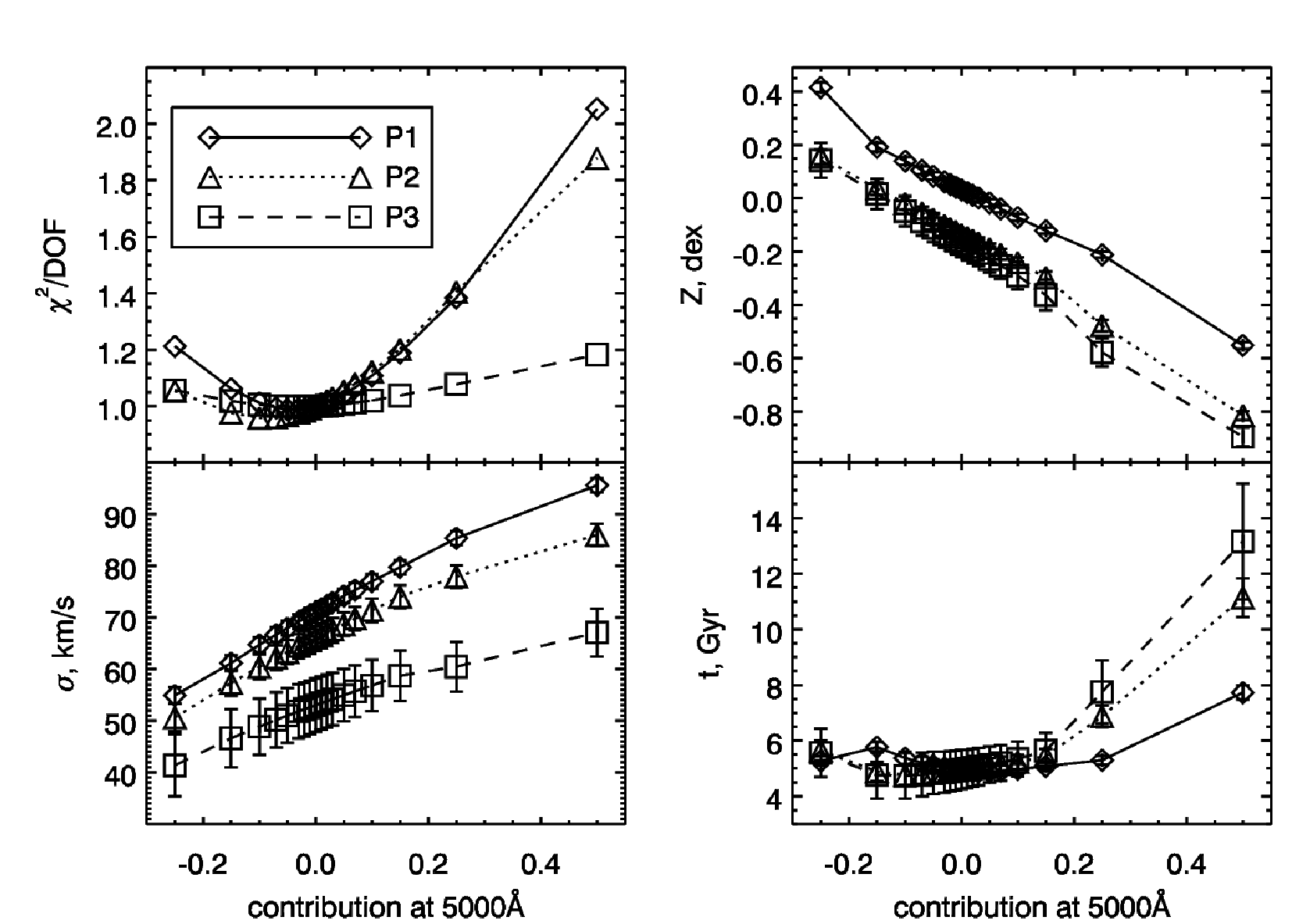}
\caption{Effects of additive terms on the results. Abscissa represents the
contribution of the constant level at $\lambda$=5000\AA. Different plotting
symbols correspond to each bin of the 3-points binning.
\label{figaddcont}
}
\end{figure}

We have conducted two series of experiments: (1) adding a constant term or (2)
the spectrum itself smoothed by a 300 pixels window to simulate the
diffuse light in the spectrograph. In every series the fraction of the additive
term was between -20 and +50 per cent to model over- and undersubtraction. The
results for the constant term as a fraction of flux at 5000\AA\ are shown in
Fig.~\ref{figaddcont}. One may notice that $\chi^2$ reaches minimum on slightly
negative (over-subtraction) values of the additive term because we did not
change flux uncertainties during our experiments. The remarkable result is the
stability of age estimations on a wide range of additive components (-25 to
15 per cent). This is an important advantage of the pixel-fitting technique over
Lick indices, more sensitive to additive continuum. Metallicity and velocity
dispersion exhibit the expected behaviour: growth of $\sigma$ and fall of Z.
Indeed within a range of contribution between -5 and 5 per cent changes are
quite small ($\sim$8 per cent for $\sigma$, and $\sim$0.1~dex for Z) though
significant.

In order to test the consequences of bad sky subtraction we made two
additional experiments: we tried to fit the data, where the sky spectrum was
represented by a low-order polynomial continuum and with no sky subtraction at
all. We excluded four regions of the spectrum containing bright emission lines:
HgI $\lambda=$4358\AA, 5461\AA, [NI] $\lambda=$ 5199\AA, and [OI] $\lambda=$
5577\AA. The experiments were made for the 3-points binning of the data,
demonstrating the effects for high, intermediate, and low surface brightness
(see Table~\ref{tabpar3bin}). Basically we found no significant difference for
the "P1"\ and "P2"\ bins between the parameters for the correct sky
subtraction and subtraction of the low-order polynomial model of the sky (see
Table~\ref{tabnsky}). "P3"\ bin gives younger age and higher metallicity, but the
estimations are in agreement with the normal sky subtraction within 2$\sigma$
However, as expected, when sky is not subtracted at all we find sizeable
bias on age, metallicity, and velocity dispersion estimations for
the "P2"\ bin, and even stronger effect for "P3". Due to additive continuum
metallicities are found to be lower, ages older, and velocity dispersions higher
than expected. These experiments demonstrate that for the surface brightness
down to $\mu_B=20$~mag~arcsec$^{-2}$ features of the night sky spectrum do not
affect the results of the pixel fitting procedure, and very rough sky
subtraction is sufficient to obtain the realistic estimations of kinematical and
stellar population parameters.

\begin{table}
\begin{tabular}{l c c c}
 & P1 & P2 & P3 \\
\hline
$v$,~km~s$^{-1}$ &  604.3$\pm$   1.0 &  606.0$\pm$   1.5 &  609.4$\pm$   3.2 \\
 &  604.4$\pm$   1.0 &  605.9$\pm$   1.6 &  610.3$\pm$   3.7 \\
\hline
$\sigma$,~km~s$^{-1}$ &   71.5$\pm$   1.5 &   64.9$\pm$   2.4 &   54.4$\pm$   5.2 \\
 &   80.5$\pm$   1.5 &   88.3$\pm$   2.2 &  105.5$\pm$   4.6 \\
\hline
t, Gyr &  4.868$\pm$ 0.210 &  4.547$\pm$ 0.310 &  3.956$\pm$ 0.731 \\
 &  4.972$\pm$ 0.185 &  7.203$\pm$ 0.390 & 12.982$\pm$ 1.641 \\
\hline
Z, dex &   0.04$\pm$  0.01 &  -0.10$\pm$  0.02 &  -0.06$\pm$  0.04 \\
 &  -0.08$\pm$  0.01 &  -0.50$\pm$  0.02 &  -0.94$\pm$  0.03 \\
\hline
\end{tabular}
\caption{Determination of the kinematical and stellar population parameters
 for a case of polynomial night sky model (first line for every parameter)
 and no sky subtraction (second line for every parameter).
\label{tabnsky}
}
\end{table}

\label{lastpage}

\end{document}